\documentclass[12pt,showpacs,aps,prd]{revtex4}
\input epsf

\begin{document}
\title{Light-cone QCD sum rules for the semileptonic decay
$\Lambda_b\rightarrow p\ell\bar\nu$}
\author{Ming-Qiu Huang and Dao-Wei Wang}
\affiliation{Department of Applied Physics, National University of Defense
Technology, Hunan 410073, China}
\date{\today}
\begin{abstract}
The exclusive semileptonic heavy baryon decay $\Lambda_b\rightarrow p\ell\bar\nu$
is investigated using light-cone sum rule method in both full QCD and HQET. The
form factors describing the decay are calculated and used to predict the decay
width and differential distribution. The total decay width obtained from full QCD
is in agreement with the previous theoretical predictions while the HQET result
are typically one order smaller. Both results are consistent with the
experimental upper limit and can be compared to the refined experimental data in
the future.
\end{abstract}
\pacs{13.30.-a, 14.20.-c, 12.39.Hg, 11.55.Hx} \maketitle
\section{Introduction}
\label{sec1}

The study of various decay and formation modes of the $b$ quark is a main source
of information about CKM matrix elements and a field for understanding
perturbative and nonperturbative QCD effects. In particular, heavy-to-light
decays are interesting because they give information on $V_{ub}$, but they are
especially difficult to calculate because of the essential presence of strong
interactions in the hadronic bound state. The form factors parameterizing the
relevant hadronic matrix elements in the heavy-to-light transitions are
nonperturbative quantities, which must be estimated in some nonperturbative
theoretical approaches. In this regards, the precise calculation of the form
factors is provided by lattice simulations. Another fruitful approach has been
the application of QCD sum rules on the light-cone \cite{BBK,LCSR,hqetsum}.

The method of light-cone sum rules (LCSR) is a hybrid of the standard technique
of QCD sum rules \`a la SVZ \cite{svzsum}, with the conventional distribution
amplitude description of the hard exclusive process \cite{HEP}. The basic idea of
SVZ sum rules is that using vacuum condensates to parameterize the nontrivial QCD
vacuum and employing the duality hypothesis to relate the experimental observable
to the theoretical calculation. Technically, operator product expansion (OPE)
based on the canonical dimension is used. The difference between SVZ sum rules
and LCSR is that the short-distance Wilson OPE in increasing dimension is
replaced by the light-cone expansion in terms of distribution amplitudes of
increasing twist.

The main nonperturbative parameter in LCSR is the distribution amplitude, also
called wave function, which corresponds to the sum of an infinite series of
operators with the same twist. In practical application the correlation function
considered in LCSR is a $T$ product inserted between physical state and vacuum,
which is to be compared to the correlation function between vacuum in the SVZ sum
rules. The main contribution to this type of correlation function comes from the
light-cone. Using vacuum condensate to characterize the QCD vacuum corresponding
to the mean field approximation, and this is valid when the momentum transfer is
not so large. When the momentum transfer becomes larger, this approximation
cannot be correct everywhere. To this end, the condensates must be replace by the
distribution amplitudes, which embodies the fast variation of the field.

Although the LCSR approach does involve a certain model dependence and the
leading-order sum rules may not be very accurate, this technique offers an
important advantage of being fully consistent with QCD perturbation theory. In
recent years there have been numerous applications of LCSR to mesons including
the determination of the form factors and the calculations of hadron matrix
elements \cite{apps,Btopi,app-h},  see \cite{LCSR,hqetsum} for a review. In
compliance with the heavy quark symmetry \cite{HQET,review}, LCSR within the
framework of the Heavy Quark Effective Theory (HQET) has also been formulated and
achieved fruitful results. The HQET light-cone sum rules have been applied to
deal with the strong couplings of heavy hadrons and heavy-to-light transitions
\cite{app-hqet}.

In this paper we extend the LCSR approach to study the exclusive semileptonic
decay $\Lambda_b\rightarrow p\ell\bar\nu$. Unlike the meson cases, the
applications of LCSR to baryons have received little attention because of the
lack of knowledge about the distribution amplitudes of higher twists. Recently a
comprehensive investigation for the baryonic distribution amplitudes has been
given in Ref. \cite{hitwist}, which makes the calculation of baryon form factors
from the LCSR approach feasible. The study for the exclusive semileptonic
$\Lambda_b$ decay into proton can be found in the literatures by using various
approaches such as the SVZ sum rules \cite{result1}, the quark model
\cite{result2} and perturbative QCD factorization theorems \cite{pqcd}. The
existing theoretical predictions vary from each other, and can differ even by
orders of magnitude. Here we shall investigate the $\Lambda_b\rightarrow
p\ell\bar\nu$ decay in LCSR and calculate the form factors of the decay within
both the framework of full QCD and HQET.

The paper is organized as follows. The full QCD LCSR is derived in Sec.
\ref{sec21}, while the LCSR in HQET is given in Sec. \ref{sec22}. Sec. \ref{sec3}
is devoted to the numerical analysis. And in Sec. \ref{sec4} the decay
distributions and widths are discussed. Finally, Sec. \ref{sec5} is our
conclusion. For completeness and convenience we list the nucleon distribution
amplitudes in the Appendix.
\section{$\Lambda_b\to p\ell\bar\nu$ decay form factors from light-cone sum rules}
\subsection{LCSR analysis in the full QCD}\label{sec21}

\subsubsection{state of the art: leading twist}

In the following LCSR analysis we adopt the current below to interpolate the
$\Lambda_b$ baryon state
\begin{equation}
j_\Lambda=\epsilon_{ijk}(u^i C\rlap/z d^j)\gamma_5\rlap/z b^k,
\end{equation}
where $C$ is the charge conjugation matrix, and $i$, $j$, $k$ denote the color
indices. The auxiliary four-vector $z$, which satisfies the light-cone condition
$z^2=0$, is introduced to project out the main contribution on the light-cone.
For the interpolating current used in the sum rule approach, there have been many
discussions \cite{current}. What worth noting is that the current interpolating a
given state is not unique, the practical criterion is that the coupling between
the interpolating current and the given state must be strong enough. The coupling
constant between the interpolating current and the vacuum is
\begin{equation}
\langle 0\mid j_\Lambda \mid
\Lambda_b(P')\rangle=f_{\Lambda}z\cdot P'\rlap/z\,u_\Lambda,
\label{ccons-fqcd}
\end{equation}
in which $u_\Lambda$ is the $\Lambda_b$ baryon spinor and $P'$ is the $\Lambda_b$
four-momentum.

Our analysis for the decay $\Lambda_b\rightarrow p\ell\bar\nu$ form factors is
analogous to that for the nucleon form factors \cite{nucleon}. The correlation
function we consider in this work is
\begin{equation}
\label{tnu} T_\nu(P,q)=i\int d^4xe^{iq\cdot x}\langle 0\mid
T\{j_\Lambda(0) j_\nu(x)\} \mid P\rangle
\end{equation}
where $j_ \nu=\bar b\gamma_\nu(1-\gamma_5)u$ is the weak current. The hadronic
matrix element of $j_\nu$ inserted between $\Lambda_b$ and proton state defines
the form factors
\begin{eqnarray}
\langle\Lambda_b(P-q)\mid j_\nu \mid P \rangle&=&\bar
u_\Lambda(P-q)\left[f_1\gamma_\nu-i\frac{f_2}{m_\Lambda}\sigma_{\nu\mu}q^\mu-\frac{f_3}
{m_\Lambda}q_\nu\right.\nonumber\\&-&\left.\left(g_1\gamma_\nu+i\frac{g_2}{m_\Lambda}
\sigma_{\nu\mu}q^\mu+\frac{g_3}
{m_\Lambda}q_\nu\right)\gamma_5\right]N(P),
\end{eqnarray}
in which $m_\Lambda$ is the $\Lambda_b$ mass, $N(P)$ denotes the proton spinor
and satisfy $\rlap/PN(P)=MN(P)$, where $M$ is the proton mass and $P$ is its
four-momentum. The six form factors $f_i$ and $g_i$ are functions of the momentum
transfer $q^2$. In the case of massless final leptons with $q_\mu\bar
\ell\gamma^\mu(1-\gamma_5)\nu_\ell=0$, $f_3$ and $g_3$ do not contribute. In the
following we shall not consider them.

By inserting a complete set of states the correlation function (\ref{tnu}) can be
represented as
\begin{equation}\label{tnuhadr}
z^\nu T_\nu=\frac{2f_\Lambda}{m_\Lambda^2-P'^2}(z\cdot
P')^2\left[f_1\rlap/z+f_2\frac{\rlap/z\rlap/q}{m_\Lambda}-
\left(g_1\rlap/z-g_2\frac{\rlap/z\rlap/q}{m_\Lambda}\right)\gamma_5\right]N(P)+\cdots,
\end{equation}
where $P'=P-q$ and the dots stand for the higher resonances and continuum. The
contraction with $z^\nu$ can simplify the Lorentz structure and remove the $\sim
z_\nu$ terms which give the subdominant contribution on the light-cone. The form
factors enter this expression as the residues of the contribution of ground state
$\Lambda_b$ baryon.

On the theoretical side, at large Euclidean momenta $P'^2$ and $q^2$ the
correlation function (\ref{tnu}) can be calculated in perturbation theory. The
diagram contributing to the correlation function is shown in Fig. 1.  In the
leading order of $\alpha_s$ a simple calculation gives
\begin{equation}\label{tnuth}
z^\nu T_\nu=-2(C\rlap/z)_{\alpha\beta}\rlap/z(1-\gamma_5)_\mu\int
d^4x\int\frac{d^4k}{(2\pi)^4}\frac{z\cdot k}{k^2-m_b^2}\;e^{i(k+q)\cdot
x}\;\langle 0\mid \epsilon_{ijk}u^i_\alpha(0)u^j_\mu(x) d^k_\beta(0)\mid
P\rangle\;.
\end{equation}
In the light-cone limit $x^2\rightarrow0$, the matrix element of the remaining
three quark operators sandwiched between the proton state and vacuum can be
parameterized by the leading twist distribution amplitudes \cite{HEP,hitwist,wf}
\begin{eqnarray}\label{lt}
&4&\langle 0\mid \epsilon_{ijk}u^i_\alpha(a_1x)u^j_\mu(a_2x)
d^k_\beta(a_3x)\mid
P\rangle\nonumber\\&=&\mathcal{V}_1(\rlap/PC)_{\alpha\mu}(\gamma_5N)_{\beta}+
\mathcal{A}_1(\rlap/P\gamma_5 C)_{\alpha\mu}N_{\beta}+
\mathcal{T}_1(P^\nu
i\sigma_{\lambda\nu}C)_{\alpha\mu}(\gamma^\lambda\gamma_5
N)_{\beta}.
\end{eqnarray}
Each distribution amplitudes $\mathcal{V}_1$, $\mathcal{A}_1$ and
$\mathcal{T}_1$ can be represented as Fourier integral over the longitudinal
momentum fractions $x_1$, $x_2$, $x_3$ carried by the quarks inside the nucleon
with $\Sigma_i x_i=1$,
\begin{equation}
F(a_kp\cdot x)=\int\mathcal{D}x\;e^{-ip\cdot x\Sigma _jx_ja_j}F(x_i).
\end{equation}
The integration measure is defined as
\begin{equation}
\int\mathcal{D}x=\int_0^1dx_1dx_2dx_3\delta(x_1+x_2+x_3-1).
\end{equation}
The normalization of $\mathcal{V}_1$ at the origin defines the nucleon coupling
constant $f_N$,
\begin{equation}\label{fn}
\langle 0\mid \epsilon_{ijk}[u^i(0)C\rlap/ zu^j(0)]\gamma_5\rlap/
z d^k_\beta(0)\mid P\rangle=f_N z\cdot P\rlap/ z N(P).
\end{equation}
With these definitions it is easy to obtain the leading twist
contribution to Eq. (\ref{tnuth}),
\begin{equation}
\label{tnu3} z^\nu T_\nu=-(z\cdot P)^2\int\mathcal{D}x\;
x_2\frac{V_1+A_1-2T_1}{(x_2P-q)^2-m_b^2}\rlap/z(1-\gamma_5)N(P)+\cdots,
\label{leading-result}
\end{equation}
where the ellipses stand for contributions that are nonleading in the infinite
momentum frame kinematics $P\rightarrow\infty$, $q\sim \mbox{const.}$, $z\sim
1/P$. Note that the light-cone expansion for obtaining Eq. (\ref{tnu3}) only
remains valid at small and intermediate momentum transfer square $q^2\leq
m_b^2-O(\Lambda_{\rm QCD}m_b)$. For larger $q^2$ the higher twist contribution
will be enhanced and the light-cone expansion becomes meaningless \cite{Btopi}.

In the leading twist there are only two form factors, $f_1$ and $g_1$, left.
Before proceeding we need the dispersion representation of Eq. (\ref{tnu3}) for
the later use in the matching between the QCD calculation and the hadronic
representation in Eq. (\ref{tnuhadr}),
\begin{equation}\label{disper3}
z^\nu T_\nu=(P\cdot z)^2\int_{m_b^2}^\infty ds\;
\frac{\rho(s,Q^2)}{s-P'^2}\;\rlap/z(1-\gamma_5)N(P)+\cdots,
\end{equation}
where $Q^2=-q^2$. If neglecting the terms of $\mathcal{O}(M^2/Q^2)$-which is
consistent with twist-$3$ accuracy-such a representation can be easily obtained
by substitution $(x_2P-q)^2-m_b^2\rightarrow-x_2(s-P'^2)$ with
$s=(1-x_2)Q^2/x_2+m_b^2/x_2$ in Eq. (\ref{tnu3}).

In the applications of QCD sum rule method the commonly adopted duality
assumption approximates the higher resonances and continuum contributions by the
same dispersion integral as that used in Eq. (\ref{disper3}) but with the
integration variable running above the continuum threshold $s_0$. Then the
matching between Eqs. (\ref{tnuhadr}) and (\ref{disper3}) is equivalent to
restrict the integration in the dispersion representation below the continuum
threshold. This upper bound in $s$ correspond to the lower bound in the momentum
fraction: $x_2>(Q^2+m_Q^2)/(Q^2+s_0)$. Following the standard procedure in QCD
sum rule we introduce the Borel transformation to suppress the higher mass
contributions
\begin{equation}
-\frac{1}{(x_2P-q)^2-m_b^2}=\frac{1}{x_2(s-P'^2)}\rightarrow\frac{1}{x_2}
exp\left\{-\frac{s}{M_B^2}\right\}.
\end{equation}
Equating the Borel transformed Eq. (\ref{tnuhadr}) and Eq.
(\ref{disper3}) we finally arrives at the sum rule
\begin{eqnarray}
f_1^{3}=g_1^{3} &=& \frac{1}{2f_\Lambda}\int\mathcal{D}x(V_1+A_1-2T_1)exp\left(
-\frac{x_2Q^2-(x_2m_\Lambda^2-m_b^2)}{x_2M_B^2}\right)\Theta\left(x_2-\frac{Q^2+m_b^2}
{Q^2+s_0}\right),\nonumber\\
\end{eqnarray}
and $f_2^{3}=g_2^{3}=0$, where the superscript $3$ denotes the
twist-$3$ results. The $\Theta$ function originates from the
restriction that the integration of the spectral density must lies
within the duality region $s<s_0$. In the limit
$Q^2\rightarrow\infty$ this restriction confines the integration
region to $x_2\rightarrow 1$. Then it is obvious that the main
contribution in this approximation comes from the configuration
with a $u$ quark carrying almost the total momentum. And this is
precisely the soft, or Feynman mechanism and gives subleading
contribution at very large momentum transfer. Indeed using the
asymptotic distribution amplitudes $V_1(x_i)=120f_Nx_1x_2x_3$,
$A_1(x_i)=0$ and $T_1(x_i)=V_1(x_i)$, and  expanding in powers of
$1/Q^2$ we find that the sum rule in the $Q^2\rightarrow\infty$
limit behaves as
\begin{equation}
f_1^3=-\frac{10}{Q^8}\frac{f_N}{f_\Lambda}e^{(m_\Lambda^2-m_b^2)/M_B^2}
\int_0^{s_0-m_b^2}dss^3e^{-s/M_B^2},
\end{equation}
i.e., it is suppressed by two additional powers of $1/Q^2$ compared with the
expected asymptotic behavior based on the dimension counting rule. This strong
suppression can be compensated by the contributions of the higher twist
distribution amplitudes.

\subsubsection{beyond the leading twist}

Besides the radiative corrections, the higher twist distribution amplitudes do
contribute to the correlation function in Eq. (\ref{tnu}). There are two kinds of
higher twist corrections. One is due to the corrections to the heavy quark
propagator in the back ground color field \cite{propagator}, which correspond to
the configuration in the Fock space with a gluon field or quark-antiquark pair in
addition to the three valance quarks and give rise to four-particle (and
five-particle) nucleon distribution amplitudes. Such effects are usually expected
to be small \cite{multi-part} and we would not take them into account. The other
one comes from different Lorentz structures and less singular contributions on
the light-cone in the decomposition of the matrix element of the three valance
quark operators in Eq. (\ref{tnuth}), besides those leading twist ones given in
Eq. (\ref{lt}). The detailed expression will not be written down here and can be
found in \cite{hitwist} and in the Appendix.

Taking the higher twist distribution amplitudes into account and proceeding as
what we have done for the leading twist case, we obtain the following result
\begin{eqnarray}
z^\nu T_\nu&=&-(P\cdot z)^2\left[\int dx_2\;
x_2\frac{B_0(x_2)}{k^2-m_b^2}-M^2\int dx_3\;
x_3\left(\frac{\mathcal{V}_1^{M(d)}(x_3)}{(k^2-m_b^2)^2}-
\frac{2m_b^2\mathcal{V}_1^{M(d)}(x_3)}
{(k^2-m_b^2)^3}\right)\right.\nonumber\\
&&+M^2\int dx_2\;\frac{x_2^2B_1(x_2)}{(k^2-m_b^2)^2}+2M^2\int dx_2\;
x_2\left(\frac{B_2(x_2)}{(k^2-m_b^2)^2}+\frac{2Q^2B_2(x_2)}{(k^2-m_b^2)^3}\right)
\nonumber\\
&&\left.+2M^4\int dx_2\;\frac{x_2^3B_3(x_2)}{(k^2-m_b^2)^3}\right]
\rlap/z(1-\gamma_5)N(P)
-(P\cdot z)^2\left[M\int dx_2\;\frac{x_2C_1(x_2)}{(k^2-m_b^2)^2}\right.\nonumber\\
&&\left.+2M^3\int dx_2\;\frac{x_2^2C_2(x_2)}{(k^2-m_b^2)^3}\right] \rlap/z\rlap/q
(1+\gamma_5)N(P)+\cdots,\label{ht-res}
\end{eqnarray}
where $k=x_2P-q$ or $k=x_3P-q$, depending on the integral variables, and the
functions $B_i$, $C_i$ are defined by
\begin{eqnarray}
B_0&=&\int_0^{1-x_2}dx_1(V_1+A_1-2T_1)(x_1,x_2,1-x_1-x_2),\nonumber\\
B_1&=&-2\tilde{V_1}+\tilde{V_2}+\tilde{V_3}+\tilde{V_4}
+\tilde{V_5}-2\tilde{A_1}+\tilde{A_2}-\tilde{A_3}-\tilde{A_4}
+\tilde{A_5}\nonumber\\
&&+ 4\tilde{T_1}-2\tilde{T_2}-2\tilde{T_5}-4\tilde{T_7}
-4\tilde{T_8},\nonumber\\
B_2&=&\tilde{\tilde{T_2}}-\tilde{\tilde{T_3}}-\tilde{\tilde{T_4}}
+\tilde{\tilde{T_5}}+\tilde{\tilde{T_7}}+\tilde{\tilde{T_8}} ,\nonumber\\
B_3&=&\tilde{\tilde{V_1}}-\tilde{\tilde{V_2}}
-\tilde{\tilde{V_3}}-\tilde{\tilde{V_4}}-\tilde{\tilde{V_5}}
+\tilde{\tilde{V_6}}+\tilde{\tilde{A_1}}-\tilde{\tilde{A_2}}
+\tilde{\tilde{A_3}}+\tilde{\tilde{A_4}}-\tilde{\tilde{A_5}}
+\tilde{\tilde{A_6}}\nonumber\\&& -2(-\tilde{\tilde{T_1}}+\tilde{\tilde{T_2}}
+\tilde{\tilde{T_5}}-\tilde{\tilde{T_6}}
+2\tilde{\tilde{T_7}}+2\tilde{\tilde{T_8}}),\nonumber\\
C_1&=&\tilde{V_1}-\tilde{V_2}-\tilde{V_3}
+\tilde{A_1}-\tilde{A_2}+\tilde{A_3}-2(\tilde{T_1} -\tilde{T_3}-
\tilde{T_7}),\nonumber\\
C_2&=&-\tilde{\tilde{V_1}}+\tilde{\tilde{V_2}}
+\tilde{\tilde{V_3}}+\tilde{\tilde{V_4}}+\tilde{\tilde{V_5}}
-\tilde{\tilde{V_6}}-(\tilde{\tilde{A_1}}-\tilde{\tilde{A_2}}
+\tilde{\tilde{A_3}} +\tilde{\tilde{A_4}}-\tilde{\tilde{A_5}}
+\tilde{\tilde{A_6}})\nonumber\\&&-2(-\tilde{\tilde{T_1}}
+\tilde{\tilde{T_3}}+\tilde{\tilde{T_4}}
-\tilde{\tilde{T_6}}+\tilde{\tilde{T_7}}+\tilde{\tilde{T_8}}).
\end{eqnarray}
The distribution amplitudes with tildes are defined via integrations as follow
\begin{eqnarray}
\tilde{V}(x_2)&=&\int_1^{x_2}dx'_2\int_0^{1-x'_2}dx_1V(x_1,x'_2,1-x_1-x'_2),\nonumber\\
\tilde{\tilde{V}}(x_2)&=&\int_1^{x_2}\int_1^{x'_2}dx''_2\int_0^{1-x''_2}dx_1
V(x_1,x''_2,1-x_1-x''_2).
\end{eqnarray}
They originate from the partial integration in order to get rid of the factor
$1/P\cdot x$ which appears in the distribution amplitudes. The surface terms for
each distribution amplitude sum to zero so they do not contribute. The term $B_0$
corresponds to the leading twist contribution, which is given in
(\ref{leading-result}). It is apparent that there arise the form factors $f_2$
and $g_2$ in (\ref{ht-res}) due to the higher twist contributions.

The Borel transformation and the continuum subtraction are equivalent to the
substitutions below:
\begin{eqnarray}\label{sub}
&&\int dx\;\frac{\rho(x)}{(k^2-m_b^2)^2}=\int\frac{dx}{x^2}\frac{\rho(x)}
{(s'-P'^2)^2}
\rightarrow\frac{1}{M_B^2}\int^1_{x_0}dx\;\frac{\rho(x)}{x^2}e^{-s'/M_B^2}+
\frac{e^{-s_0/M_B^2}\rho(x_0)}{Q^2+m_b^2+x_0^2M^2},\nonumber\\
&&\int dx\;
\frac{\rho(x)}{(k^2-m_b^2)^3}=-\int\frac{dx}{x^3}\frac{\rho(x)}{(s'-P'^2)^3}
\rightarrow-\frac{1}{2M_B^4}\int^1_{x_0}dx\;\frac{\rho(x)}{x^3}e^{-s'/M_B^2}
\nonumber\\&&- \frac{e^{-s_0/M_B^2}\rho(x_0)}{2M_B^2(Q^2+m_b^2+x_0^2M^2)}+
\frac{e^{-s_0/M_B^2}x_0^2}{2(Q^2+m_b^2+x_0^2M^2)}\frac{d}{dx_0}
\left(\frac{\rho(x_0)}{x_0(Q^2+m_b^2+x_0^2M^2)}\right),
\end{eqnarray}
where
\begin{equation}
s'=(1-x)M^2+\frac{m_b^2+(1-x)Q^2}{x},
\end{equation}
and $x_0$ is the positive solution of the quadratic equation for $s'=s_0$:
\begin{equation}
2M^2x_0=\sqrt{(Q^2+s_0-M^2)^2+4M^2(Q^2+m_b^2)}-(Q^2+s_0-M^2).
\end{equation}
Those terms we do not write explicitly in Eq. (\ref{sub}) give no contributions
since all the $\rho(x)$ and the corresponding first order derivative vanishing at
$x=1$.

Putting all the results together, it is straightforward to get the final sum
rules:
\begin{eqnarray}\label{sr-f1}
&&-2f_\Lambda f_1 e^{-m_\Lambda^2/M_B^2}=-\int_{x_0}^1dx_2\;e^{-s'/M_B^2}\left\{
B_0+\frac{M^2}{M_B^2}\left[\frac{\mathcal{V}_1^{M(d)}(x_2)}{x_2}+
\frac{m_b^2}{M_B^2}\frac{\mathcal{V}_1^{M(d)}(x_2)}{x_2^2}\right.\right.\nonumber\\&&
-B_1(x_2)-2\left.\left.\frac{B_2(x_2)}{x_2}-2\frac{Q^2}{M_B^2}\frac{B_2(x_2)}{x_2^2}
+\frac{M^2}{M_B^2}B_3(x_2)\right]\right\}+\frac{M^2e^{-s_0/M_B^2}}{m_b^2+Q^2+x_0^2
M^2}\nonumber\\&&\left[-x_0\mathcal{V}_1^{M(d)}(x_0)-\frac{m_b^2}{M_B^2}
\mathcal{V}_1^{M(d)}(x_0) +x_0^2B_1(x_0)+2x_0B_2(x_0)-2\frac{Q^2}{M_B^2}B_2(x_0)
-\frac{M^2}{M_B^2}x_0^2B_3(x_0)\right]\nonumber\\&&+\frac{M^2e^{-s_0/M_B^2}x_0^2}
{m_b^2+Q^2+x_0^2M^2}
\frac{d}{dx_0}\left(\frac{m_b^2\mathcal{V}_1^{M(d)}(x_0)+2Q^2B_2(x_0)+M^2
x_0^2B_3(x_0) }{m_b^2+Q^2+x_0^2M^2} \right) ,
\end{eqnarray}
and
\begin{eqnarray}\label{sr-f2}
&&-2\frac{f_\Lambda f_2}{Mm_\Lambda}
e^{-m_\Lambda^2/M_B^2}=\frac{1}{M_B^2}\int_{x_0}^1\frac{dx_2}{x_2}\;e^{-s'/M_B^2}\left(
C_1(x_2)-\frac{M^2}{M_B^2}C_2(x_2)\right)+\frac{x_0e^{-s_0/M_B^2}}{m_b^2+Q^2+x_0^2M^2}
\nonumber\\&&\left(C_1(x_2)-\frac{M^2}{M_B^2}C_2(x_2)\right)+
\frac{M^2e^{-s_0/M_B^2}x_0^2}{m_b^2+Q^2+x_0^2M^2}
\frac{d}{dx}\left(\frac{x_0C_2(x_0)}{m_b^2+Q^2+x_0^2M^2}\right).
\end{eqnarray}
For the sum rules for the form factors $g_1$ and $g_2$ are identical with those
for the $f_1$ and $f_2$, $f_1=g_1$ and $f_2=g_2$, we will only discuss the
results for $f_1$ and $f_2$ in the following sections.

In the above sum rules the four form factors are all functions of the
distribution amplitudes. Substituting into the asymptotic distribution amplitudes
and expanding in $1/Q^2$ we can get the asymptotic behavior of the form factors
in the limit $Q^2\rightarrow\infty$ as
\begin{eqnarray}
-f_\Lambda f_1 e^{-(m_\Lambda^2-m_b^2)/M_B^2}&=&\frac{M^2}{M_B^2}
\frac{f_N}{Q^6}\int_0^{s_0-m_b^2}ds\;
e^{-s/M_B^2}\left[M_B^2\left(\frac{37}{3}+2\frac{\lambda_1}{f_N}\right)s
-\frac{1}{3}\frac{\lambda_2}{f_N}s^2\right],
\nonumber\\
-\frac{f_\Lambda f_2}{Mm_\Lambda}
e^{-(m_\Lambda^2-m_b^2)/M_B^2}&=&\frac{f_N}{Q^8} \int_0^{s_0-m_b^2}ds \;
e^{-s/M_B^2}\left[M^2(11+\frac{8}{5}\frac{
\lambda_1}{f_N})s+(5+3\frac{\lambda_1}{f_N})s^2\right].\label{sr-asy}
\end{eqnarray}
In this limit the order $\mathcal{O}(x^2)$ correction which is proportional to
$\mathcal{V}_1^{M(d)}$ goes as $1/Q^8$ and gives no contribution to the $f_1$ sum
rule. It should be noted that we are working with the soft contribution to the
form factors. If the hard contribution, which acts as a part of the radiative
correction, is taken into account the true asymptotic behavior of $f_1$ is
expected to be $\sim1/Q^4$, based on the case study of the pion form factors
\cite{pion}.

Unlike the standard QCD sum rule approach, LCSR behaves well under the heavy
quark limit. Taking the heavy quark limit in the sum rules is equivalent to the
following substitutions \cite{app-h,H-limit}:
\begin{equation}
M_B^2\rightarrow 2m_b\tau,\hspace{0.3cm}s_0\rightarrow m_b^2+2m_b\omega_0
\end{equation}
Together with the asymptotic distribution amplitudes we obtain the
sum rules at the origin in the heavy quark limit
\begin{eqnarray}
-2f_\Lambda f_1(0)
e^{-\bar\Lambda/\tau}&=&\frac{16}{3}\frac{f_N}{m_b^4}
\left[60\int_0^{\omega_0}ds
s^3e^{-s/\tau}-M^2\frac{\lambda_1}{f_N} \int_0^{\omega_0}ds
se^{-s/\tau}\right],\nonumber\\
-2f_\Lambda f_2(0) e^{-\bar\Lambda/\tau}& =& M
\frac{16}{9}\frac{f_N}{m_b^3}(5+3\frac{\lambda_1}{f_N})
\int_0^{\omega_0}ds s^3e^{-s/\tau},
\end{eqnarray}
where the effective mass $\bar\Lambda$ is defined as $\bar\Lambda=m_\Lambda-m_b$.
\subsection{LCSR in the HQET} \label{sec22}

In the HQET we use the following current in the application of LCSR,
\begin{equation}
j_v=\epsilon_{ijk}(u^i C\rlap/z d^j)\gamma_5\rlap/z h_v^k,
\end{equation}
where the heavy quark field is denoted by $h_v$ and $v$ is the heavy quark
velocity. The corresponding coupling constant is defined as
\begin{equation}
\langle 0\mid j_v \mid \Lambda_b(v)\rangle=\hat{f}_{\Lambda}z\cdot
v\rlap/z\,u_\Lambda(v),
\end{equation}
in which $u_\Lambda$ is the $\Lambda_b$ baryon spinor in the HQET. The two form
factors for the decay $\Lambda_b\rightarrow p\ell\bar\nu$ in the HQET are given
by the weak matrix element \cite{hform}
\begin{equation}
\langle\Lambda_b(v)\mid j_\nu \mid P \rangle=\bar
u_\Lambda(v)\gamma_\nu(1-\gamma_5)(F_1+F_2\rlap/v)N(P).
\end{equation}
Both $F_1$ and $F_2$ are functions of the momentum transfer $\omega=v\cdot P$.
The correlation function we consider in this case is
\begin{equation}
\label{tnuh} T_\nu(P,q)=i\int d^4xe^{iq\cdot x}\langle 0\mid
T\{j_v(0) j_\nu(x)\} \mid P\rangle.
\end{equation}
Then by inserting a complete set of states we obtain the hadronic part of the
correlation function as what we have done in order to get the Eq.
(\ref{tnuhadr}),
\begin{equation}\label{tnuhadrh}
z^\nu T_\nu=\frac{\hat{f}_\Lambda}{m_b}\frac{z\cdot v z\cdot
P'}{\bar\Lambda-\omega'}\rlap/z(1-\gamma_5)(F_1+F_2\rlap/v)N(P)+\cdots,
\end{equation}
where the variable $\omega'$ is defined as $\omega'=v\cdot (P-q)=v\cdot P'$. It
should be remembered that the relevant variable for the dispersion relation is
$\omega'$ and the consequent Borel transformation is performed on it, too.

The theoretical part can be calculated straightforwardly,
\begin{eqnarray}\label{zth}
z^\nu T_\nu&=&z\cdot v z\cdot P\left[\int
\frac{\mathcal{D}x}{s'-\omega'}\left[\frac{K_1}{2}-
\frac{M^2}{4\omega^2}K_2\right]+\frac{M^2}{4}\int
dx_3\frac{\mathcal{V}_1^{M(d)}(x_3)}
{(s'-\omega')^3}\right]\rlap/z(1-\gamma_5)N(P) \nonumber\\&+&
z\cdot v z\cdot P \frac{M}{4\omega}\int\frac{\mathcal{D}x}
{s'-\omega'}H_1\rlap/z\rlap/v (1+\gamma_5)N(P),
\end{eqnarray}
where $s'=(1-x)\omega$ and the substitution $(xP-q)\cdot v=-(s'-\omega')$ has
been made. The functions $K_i$ ($i=1,2$) and $H_1$ are defined by
\begin{eqnarray}
K_1&=&V_1+A_1-2T_1,\nonumber\\
K_2&=&T_2-T_3-T_4+T_5+T_7+T_8,\nonumber\\
H_1&=&-V_1+V_2+V_3-A_1+A_2-A_3+2(T_1-T_3-T_7).
\end{eqnarray}
The Borel transformation and continuum subtraction in HQET are performed through
substitutions analogous to Eq. (\ref{sub}),
\begin{eqnarray}
\int dx\;\frac{\rho(x)}{s'-\omega'}& \rightarrow&\int
dx\rho(x)e^{-s'/T}{\Theta}(\omega_c-s'),\nonumber\\
\int dx\;\frac{\rho(x)}{(s'-\omega')^3}& \rightarrow&\frac{1}{2 T^2}\int
dx\rho(x)e^{-s'/T}{\Theta}(\omega_c-s')\nonumber+
\frac{e^{-\omega_c/T}\rho(x_c)}{2T\omega}-
\frac{e^{-\omega_c/T}\rho'(x_c)}{2\omega^2},
\end{eqnarray}
in which $x_c=max(0,1-\omega_c/\omega)$ and the prime on $\rho$ denotes the
derivative. Equating the Borel transformed hadronic part with the theoretical
part of the correlation function thus substituted, we arrive at the final sum
rules,
\begin{eqnarray}
\frac{\hat{f}_\Lambda}{m_b} F_1(\omega)e^{-\bar\Lambda/T}&=&\int\mathcal{D}x
e^{-s'/T}\left(\frac{1}{2}K_1 -\frac{M^2}{4\omega^2}K_2\right)\Theta(\omega_c-s')
+\frac{M^2}{8T^2}\int_{x_c}^1 dx_3 \mathcal{V}_1^{M(d)}e^{-s'/T}
\nonumber\\&+&\frac{M^2}{8T\omega} \mathcal{V}_1^{M(d)}(x_c)
e^{-\omega_c/T}-\frac{M^2}{8\omega^2}
\frac{d}{dx_c}\mathcal{V}_1^{M(d)}(x_c)e^{-\omega_c/T} \label{hf1}\end{eqnarray}
and
\begin{eqnarray} \frac{\hat{f}_\Lambda}{m_b}
F_2(\omega)e^{-\bar\Lambda/T}&=&\frac{M}{4\omega}\int\mathcal{D}x
H_1e^{-s'/T}{\Theta}(\omega_c-s').\label{hf2}
\end{eqnarray}
The $K_1$ term in Eq. (\ref{hf1}) corresponds to the leading twist contribution
to the form factor. For the above given sum rules it is obvious that the very
soft proton will enhance the higher twist contribution and invalidate the
expansion. In practice we will stay away from this region.

For the asymptotic behavior of the form factors in the limit
$\omega\rightarrow\infty$, we have
\begin{eqnarray}
\frac{\hat{f}_\Lambda}{m_b} F_1(\omega)e^{-\bar\Lambda/T}&=&
\frac{f_N}{\omega^4}\left[\frac{M^2}{12}\left(37+6\frac{\lambda_1}{f_N}+
2\frac{\lambda_2}{f_N}\right)\delta_1(\omega_c/T)-
\frac{5}{3}\delta_3(\omega_c/T)\right],\nonumber\\
\frac{\hat{f}_\Lambda}{m_b}
F_2(\omega)e^{-\bar\Lambda/T}&=&\frac{f_N}{\omega^3}
\left(5+3\frac{\lambda_1}{f_N}\right)
\delta_2(\omega_c/T),\label{hsr-asy}
\end{eqnarray}
where the functions $\delta_n$ are defined as
\begin{equation}
\delta_n(\omega_c/T)=\frac{1}{n!}\int_0^{\omega_c} ds s^ne^{-s/T}
\end{equation}
and the $\delta_3$ term corresponds to the leading twist contribution.

\section{numerical analysis}\label{sec3}

In the numerical evaluation, the input parameters can be classified into two
parts. One part belongs to the nucleon side, which have been determined before.
The other part goes to the $\Lambda_b$ baryon sector, which will be given in the
following part. Since there have been detailed analysis and discussion about the
parameters on the nucleon side in \cite{hitwist}, we will not dwell on the
specific details, but only list the numerical values that will be used in the
following analysis,
\begin{eqnarray}
f_N=5.3\mathcal{\Theta}10^{-3}\mbox{GeV}^2,\hspace{0.3cm} \lambda_1=-2.7
\mathcal{\Theta}10^{-2} \mbox{GeV}^2,\hspace{0.3cm} \lambda_2=5.1\mathcal{\Theta}
10^{-2}\mbox{GeV}^2.
\end{eqnarray}

In the full QCD numerical analysis, we need the coupling constant defined by Eq.
(\ref{ccons-fqcd}). In order to get an estimate, we use QCD sum rule method and
consider the correlation function
\begin{equation}
\label{cf-corr} \Pi(q^2)=i\int d^4xe^{iq\cdot x}\langle 0\mid
T\{j_\Lambda(x) \bar j_\Lambda(0)\} \mid 0\rangle.
\end{equation}
The hadronic representation can be obtained immediately
\begin{equation}
\Pi=\frac{2f_\Lambda^2}{m_\Lambda^2-q^2}(z\cdot
q)^3\rlap/z+\cdots.
\end{equation}
On the other hand, the correlation function (\ref{cf-corr}) can be calculated
perturbatively in the deep Euclidean region. Then making use of the usual duality
assumption and employing Borel transformation to suppress continuum contributions
we arrive at the sum rule to dimension $6$ in the OPE
\begin{equation}
2f_\Lambda^2e^{-m_\Lambda^2/M_B^2}=\int_{{m_b^2}}^{s_0}
ds\rho(s)e^{-s/M_B^2}, \label{f-sr}
\end{equation}
where the spectral density is
\begin{equation}
\rho(s)=\frac{s}{2^6\cdot5\pi^4}\left(1-\frac{m_b^2}{s}\right)^5-
\langle\frac{\alpha_s}{\pi}G^2\rangle\frac{m_b^2}
{2^5\cdot3\pi^2s^2} \left(1-\frac{m_b^2}{s}\right)\left(1-\frac
{2m_b^2}{s}\right).
\end{equation}
At the working window $38<s_0<40\mbox{GeV}^2$ and $1<M_B^2<3\mbox{GeV}^2$ the
numerical value for the coupling constant reads
$f_\Lambda=(5\pm2)\mathcal{\Theta} 10^{-3}\mbox{GeV}^2$ from (\ref{f-sr}), where
the standard value $\langle \frac{\alpha_s}{\pi}G^2 \rangle=0.012~\mbox{GeV}^2$
is adopted.

In the numerical calculations for the form factors, $f_1$ and $f_2$, the center
values we take for the heavy quark mass and the continuum threshold are
$m_b=4.8~\mbox{GeV}$ \cite{mb-lb} and $s_0=39~\mbox{GeV}^2$ \cite{excited-mass},
and the $\Lambda_b$ baryon mass can be found in \cite{PDG2002}. Then substitute
into the explicit form of $f_\Lambda$ and vary the continuum threshold in the
range $s_0=38-40~\mbox{GeV}^2$, we find that the stability is acceptable in
$M_B^2=6-9~\mbox{GeV}^2$ for the Borel parameter. The $M_B^2$ and the $q^2$
dependence for the form factors are shown in Figs. 2 and 3, respectively.

Both the form factors in a subrange of the whole kinematical region, which
depends on the heavy quark mass, can be fitted well by the dipole formula
\begin{equation}
f_i(q^2)=\frac{f_i(0)}{a_2(q^2/m_\Lambda^2)^2+a_1q^2/m_\Lambda^2+1}.
\label{dipole}
\end{equation}
But the whole kinematical fit is not satisfactory. Below
corresponding to the asymptotic and QCD sum rule obtained
distribution amplitudes we give the coefficients for a special set
of values in Table \ref{di-fit}, for which the largest uncertainty
occurs in the end-point region $q^2\sim16~\mbox{GeV}^2$ and is
less than $10\%$.
\begin{table}[htb]
\begin{tabular}{|@{\hspace{1ex}}c|@{\hspace{1ex}}*{3}{r@{.}l
}@{\hspace{1ex}}|@{\hspace{1ex}}*{3}{r@{.}l}@{\hspace{1ex}}|}\hline
&\multicolumn{6}{c|@{\hspace{1ex}}}{asymptotic}&\multicolumn{6}{c|}{QCD sum rule}\\
\hline &\multicolumn{2}{c}{$a_2$}&\multicolumn{2}{c}{$a_1$}
&\multicolumn{2}{c|@{\hspace{1ex}}}{$f_i(0)$}&\multicolumn{2}{c}{$a_2$}
&\multicolumn{2}{c}{$a_1$}&\multicolumn{2}{c|}
{$f_i(0)$}\\
$f_1$&$2$&$381$&$-2$&$888$&$-0$&$037$&$5$&$590$&$-2$&$759$&$0$&$018$ \\
$f_2$&$2$&$582$&$-3$&$026$&$0$&$027$&$2$&$000$&$-2$&$603$&$0$&$159$ \\
\hline
\end{tabular}
\caption{The dipole fit for the form factors $f_1$ and $f_2$ in
the subregion $0<q^2<16~\mbox{GeV}^2$ with $M_B^2=8~\mbox{GeV}^2$,
$s_0=39~\mbox{GeV}^2$ and $m_b=4.8~\mbox{GeV}$.} \label{di-fit}
\end{table}

As already mentioned above in section \ref{sec21}, the light-cone expansion is
expected to hold only at $q^2\leq m_b^2-O(\Lambda_{\rm QCD}m_b)\approx 17$  ${\rm
GeV}^2$. In fact, the fast increase of the form factors in Figs. 2 and 3 near the
end-point region indicates that when the momentum transfer is large or $q^2$
approaches the kinematical limit, the dispersion represented integral in
(\ref{tnu3}) tends to grow strongly and cannot be enough suppressed by Borel
transformation in the subsequently obtained sum rules. This can also be
substantiated by the dipole fit procedure: If the formula (\ref{dipole}) is used
for the fit, then there exists no satisfactory fit for the form factors in the
whole range of the kinematical region and the uncertainty near the end-point
region is large, which can even amount to $80\%$. Taking into account this fact,
we are satisfied to fit the form factors in a subregion of the whole kinematical
region, just as what we have done in Table \ref{di-fit}, and then extrapolate to
get the behavior near the end-point.

The numerical analysis in the HQET goes parallel with that in the full QCD. What
we consider first is sum rule for the coupling constant, and the vacuum-to-vacuum
correlation function is
\begin{equation}
\label{cf-corr-hqet} \Pi(q^2)=i\int d^4xe^{iq\cdot x}\langle 0\mid
T\{j_v(x) \bar j_v(0)\} \mid 0\rangle.
\end{equation}
The phenomenological part can be obtained by inserting a complete set of states
\begin{equation}
\Pi=\frac{\hat{f}_\Lambda^2}{\bar\Lambda-\omega}(z\cdot v
)^3\rlap/z+\cdots,
\end{equation}
where $\omega=v\cdot q$. The theoretical calculation is straightforward. After
the Borel transformation and continuum subtraction we get the final sum rule
\begin{equation}
\hat{f}_\Lambda^2e^{-\bar\Lambda/T}=\frac{1}{10\pi^4}\int_0^{\omega_c}
d\omega\omega^5e^{-\omega/T}+\frac{T^2}{3\cdot2^4\pi^2}\langle
\frac{\alpha_s}{\pi}G^2 \rangle.
\end{equation}
Vary the parameters in the range $0.1<T<0.5\mbox{GeV}$,
$1.6<\omega_c<1.8\mbox{GeV}$, we arrive at the following value for the coupling
constant in the HQET $\hat{f}_\Lambda=(2.9\pm1.0)
\mathcal{\Theta}10^{-2}\mbox{GeV}^3$, where the effective mass is taken to be
$\bar\Lambda=0.8\mbox{GeV}$.

As what have been done in the full QCD, in the HQET analysis we substitute the
explicit form for $\hat{f}_\Lambda$ into the numerical analysis. We work with the
following parameters for the continuum threshold $\omega_c$ and the Borel
parameter $T$ in the HQET light-cone sum rules
\begin{equation}
\omega_c=1.6-1.8\mbox{GeV},\hspace{0.3cm} T=0.8-1.2\mbox{GeV}.
\end{equation}
The stability in the above given region is satisfactory. In Figs. 4 and 5 we show
the dependence of $F_1$ and $F_2$ on Borel parameter $T$ and momentum transfer
$\omega$. The overall behavior of the form factors $F_1$ and $F_2$ cannot be
fitted well by simple functions, but for large $\omega$, say $\omega>\omega_c$,
the following formula fits them well,
\begin{equation}
F_i(\omega)=a_0+a_1\omega+a_2\omega^2. \label{inver-dip}
\end{equation}
This is due to the fact that the higher twist contributions will
be enhanced by the small $\omega$, just as mentioned above in Sec.
\ref{sec22}. What given in Table \ref{hdi-fit} is our fit for the
center value of the effective mass $\bar\Lambda=0.8\mbox{GeV}$,
corresponding to the asymptotic and QCD sum rule obtained
distribution amplitudes.
\begin{table}[htb]
\begin{tabular}{|@{\hspace{1ex}}c|@{\hspace{1ex}}*{3}{r@{.}l
}@{\hspace{1ex}}|@{\hspace{1ex}}*{3}{r@{.}l}@{\hspace{1ex}}|}\hline
&\multicolumn{6}{c|@{\hspace{1ex}}}{asymptotic}&\multicolumn{6}{c|}{QCD sum rule}\\
\hline &\multicolumn{2}{c}{$a_2$}&\multicolumn{2}{c}{$a_1$}&
\multicolumn{2}{c|@{\hspace{1ex}}}{$a_0$}&\multicolumn{2}{c}{$a_2$}
&\multicolumn{2}{c}{$a_1$}&\multicolumn{2}{c|}
{$a_0$}\\
$F_1$&$-0$&$073$&$0$&$471$&$-0$&$802$&$-0$&$141$&$0$&
$802$&$-1$&$142$\\
$F_2$&$-0$&$003$&$0$&$044$&$-0$&$124$&$0$&$054$&$-0$&$238$&$0$&
$153$\\
\hline
\end{tabular}
\caption{The fit performed in the region $\omega>2.0\mbox{GeV}$ for the form
factors $F_1$ and $F_2$ with $\bar\Lambda=0.8\mbox{GeV}$,
$\omega_c=1.7\mbox{GeV}$ and $T=1\mbox{GeV}$.} \label{hdi-fit}
\end{table}

Based on similar reasoning, it is obvious that the bizarre behavior of the form
factors in the small $\omega$ region is due to the not enough suppression
$1/(x_2P-q)\cdot v$ in (\ref{zth}) when $\omega$ is not large enough. In order to
dodge this region we just follow the same tactic as used in the full QCD case and
fit the form factors for the large $\omega$ region, which is exactly what we have
done above.

Due to the numerically large values $\lambda_1/f_N=-5.1\pm 1.7$
and $\lambda_2/f_N=9.6\pm 3.6$, it is expected that the twist-$4$
contribution will be large as compared to the leading twist-$3$
contribution. We find that it indeed is the case in our
calculation, which can be seen more clearly in the asymptotic sum
rules (\ref{sr-asy}) and (\ref{hsr-asy}). In the full QCD, the
twist-$4$ and twist-$5$ contributions amount to approximately
$30\%$ and $80\%$ for the form factor $f_1$, respectively; while
for the form factor $f_2$, the twist-$4$ contribution is almost
the same in magnitude as the leading twist-$3$ one, at the same
time the twist-$5$ contribution is small, less than $1\%$, thus
negligible. The case is similar for both the QCD sum rule obtained
and asymptotic distribution amplitudes. When approaching the
end-point $q^2\sim22\mbox{GeV}^2$, the twist-$4$ and $5$
contributions for $f_1$ increase, but they have opposite sings and
thus tend to cancel each other and the net contribution is only
$\sim30\%$ or $\sim100\%$ to that of twist-$3$ distribution
amplitudes, corresponding to the asymptotic or sum rule obtained
distribution amplitudes; for $f_2$, the higher twist contributions
decrease and are negligible near the end-point region. The HQET
case is not so good. For $F_1$, when $\omega$ is small, the
twist-$5$ contribution is small and well under control. But when
$\omega$ gets larger, the twist-$5$ contribution tends to surpass
the leading twist-$3$ part and becomes the dominant one for the
sum rule obtained distribution amplitudes when the Borel parameter
$T$ is not large enough, while for the asymptotic distribution
amplitudes, the twist-$5$ contribution remains small, $\sim 20\%$,
in the whole kinematical region. The twist-$4$ contribution is
small for two kind distribution amplitudes, less than $10\%$, and
the twist-$6$ contribution is zero. However, the twist-$4$
contribution is expected to be large and the combination of the
twist-$3$ and $4$ contributions should be the dominant one, thus
the smallness of the twist-$4$ contribution, which results from
the missing of the twist-$4$ distribution amplitudes in the $F_1$
sum rule, is not acceptable. But as a compromise, the twist-$5$
contribution is also small and the leading twist-$3$ contribution
is the dominant one. For $F_2$, the twist-$4$ contribution is
larger than the twist-$3$ one. And there is no twist-$5$
contribution for $F_1$ and twist-$6$ contribution for both $F_1$
and $F_2$. If we take the sum of twist-$3$ and $4$ contributions
as the dominant one, then the higher twist contribution is exactly
zero in our calculation. What should be noted is that the
inclusion of twist-$6$ contribution in our calculation is not
complete, the only one included serves as an estimate of order.
For all the cases it is small, less than $1\%$, and can be
neglected. This may be seen as an indication that our results are
not sensitive to the still higher twist contributions. Besides,
the main uncertainty in our numerical analysis comes from the
uncertainties in $f_N$, $\lambda_1$ and $\lambda_2$, this can lead
to a $\sim 30\%$ uncertainty for the form factors.

\section{Semileptonic decay distributions and widths}\label{sec4}

For the decay $\Lambda_b\rightarrow p\ell\bar\nu$ in the full QCD, the
kinematical region is $0<q^2<(m_\Lambda-M)^2$, which is available from the LCSR
in this work, though the end-point behavior may be unreliable. For this end-point
region is marginal and thus has moderate effect for the total decay width. The
differential decay rate can be expressed as
\begin{eqnarray}
\frac{d\Gamma}{dq^2}&=&\frac{G_F^2|V_{ub}|^2}{192\pi^3m_\Lambda^5}
q^2\sqrt{q_+^2q_-^2} [-6f_1f_2m_\Lambda m_+q_-^2+6g_1g_2m_\Lambda
m_-q_+^2\nonumber\\&+&f_1^2m_\Lambda^2[\frac{m_+^2m_-^2}{q^2}+m_-^2-2(q^2
+2m_\Lambda M)]+g_1^2m_\Lambda^2
[\frac{m_+^2m_-^2}{q^2}+m_+^2-2(q^2 - 2m_\Lambda M)]\nonumber\\&-&
g_2^2[-2m_+^2 m_-^2+m_-^2q^2+q^2(q^2- 4Mm_\Lambda)]- f_2^2[-2m_+^2
m_-^2+m_+^2q^2+q^2(q^2+4Mm_\Lambda)]]
\end{eqnarray}
where $m_{\pm}=m_\Lambda\pm M$ and $q^2_{\pm}=q^2-m^2_{\pm}$. Using the form
factors given in Eqs. (\ref{sr-f1}) and (\ref{sr-f2}), and the $\Lambda_b$ and
proton masses \cite{PDG2002}, we can compute the differential decay rate and the
total decay width. The differential decay rate is shown in Fig. 6. The total
decay widths after the integration over the whole range of $q^2$ are given in
Table \ref{tdw-fqcd}, corresponding to the asymptotic and sum rule obtained
distribution amplitudes. The effect of the heavy quark mass is also taken into
account here. There is no apparent distinction between the two kind of results,
and within each kind the dependence on the heavy quark mass is mild. The referred
error is due to the variation of the Borel parameter and the continuum threshold
only. What should be remarked here is that the value $m_b=4.7\mbox{GeV}$ lies on
the margin of the possible region in which the whole range of the kinematical
$q^2$ can be accessed. Due to the radical increase of the form factors near the
end-point, one may suspects the reliability of the sum rule prediction in that
region, just as mentioned above in Sec. \ref{sec3}. So correspondingly, the total
decay widths using the fitted form factors are also given in Table
\ref{tdw-fqcd}. The difference is not drastic. Our results given in Table
\ref{tdw-fqcd} are in agreement with the QCD sum rule predictions both in the
full QCD and HQET \cite{result1}.

\begin{table}[htb]
\begin{tabular}{|@{\hspace{1ex}}c|*{3}{@{\hspace{1ex}}c}|*{3}{@{\hspace{1ex}}c}|}\hline
&\multicolumn{3}{c|@{\hspace{1ex}}}{asymptotic}&\multicolumn{3}{c|}{QCD sum rule}\\
\hline $m_b(\mbox{GeV})$&$4.7$&$4.8$&$4.9$&$4.7$&$4.8$&$4.9$\\
\hline $\Gamma$&$1.5\pm0.7$&$1.0\pm0.4$&$0.7\pm0.2$&$3.7\pm2.2$
&$3.1\pm1.8$&$2.6\pm1.4$\\ \hline
$\Gamma_f$&$0.56\pm0.14$&$0.44\pm0.09$&$0.35\pm0.06$&$2.4\pm1.3$
&$2.0\pm1.0$&$1.7\pm0.8$\\ \hline
\end{tabular}
\caption{Decay widths $(\mbox{in}\,
10^{-11}\times|V_{ub}|^2\,\mbox{\mbox{GeV}})$ for the semileptonic
decays $\Lambda_b\rightarrow p\ell\bar\nu$. $\Gamma$ is obtained
by integrating in the whole kinematical region using sum rule
data, while $\Gamma_f$ using the $q^2<16\mbox{GeV}^2$ fitted form
factors extrapolated to the whole region. The error only reflects
the variation of the Borel parameter and the continuum threshold
between $6<M_B^2<9\mbox{GeV}^2$ and $38<s_0<40\mbox{GeV}^2$.}
\label{tdw-fqcd}
\end{table}

The differential decay rate for the decay $\Lambda_b\rightarrow p\ell\bar\nu$ in
HQET is
\begin{eqnarray}
\frac{d\Gamma}{d\omega}&=&\frac{G_F^2|V_{ub}|^2}{12\pi^3}
[F_1^2(3m_\Lambda^2\omega+3M^2\omega-2m_\Lambda(M^2+2\omega^2)) +2
F_1F_2M(3m_\Lambda^2+M^2\nonumber\\&-&6m_\Lambda\omega+ 2\omega^2)
+F_2^2(3m_\Lambda^2\omega-M^2\omega+4\omega^3+2m_\Lambda(M^2-4\omega^2))]
\sqrt{\omega^2-M^2},
\end{eqnarray}
where the kinematical region for $\omega$ is
$M<\omega<\omega_{max}=(m_\Lambda^2+M^2)/2m_\Lambda$. Taking into the form
factors given by sum rules in Eqs. (\ref{hf1}) and (\ref{hf2}) we can obtain the
differential decay rate $d\Gamma/d\omega$ and the total decay width $\Gamma$. The
differential decay rate is shown in Fig. 7 and the total decay widths are given
in Table \ref{tdw-hqet}. Based on similar consideration, the total decay widths
corresponding to the fitted form factors are also given in Table \ref{tdw-hqet}.
The results are almost the same. Compared to the full QCD result, the decay width
in HQET is typically one order smaller for the sum rule obtained distribution
amplitudes. As for the asymptotic results, the decay width is almost the same in
magnitude. Our QCD sum rule obtained results in the HQET are comparable to those
in \cite{result2}.
\begin{table}[htb]
\begin{tabular}{|@{\hspace{1ex}}c|*{3}{@{\hspace{1ex}}c}|*{3}{@{\hspace{1ex}}c}|}\hline
&\multicolumn{3}{c|@{\hspace{1ex}}}{asymptotic}&\multicolumn{3}{c|}{QCD sum rule}\\
\hline $\bar\Lambda(\mbox{GeV})$&$0.7$&$0.8$&$0.9$&$0.7$&$0.8$&$0.9$\\
\hline $\Gamma$&$2.8\pm0.7$&$3.0\pm0.8$&$3.2\pm0.9$&$2.9\pm1.0$
&$3.1\pm1.1$&$3.4\pm1.3$\\ \hline
$\Gamma_f$&$3.5\pm0.7$&$3.7\pm0.8$&$4.0\pm0.9$&$3.4\pm0.5$
&$3.6\pm0.6$&$3.9\pm0.8$\\
\hline
\end{tabular}
\caption{Decay widths $(\mbox{in}\, 10^{-12}\times|V_{ub}|^2\,\mbox{\mbox{GeV}})$
for the semileptonic decays $\Lambda_b\rightarrow p\ell\bar\nu$ in HQET. $\Gamma$
denotes the total decay width using sum rules for $F_1$ and $F_2$, and $\Gamma_f$
using the form factors fitted in $\omega>2.0\mbox{GeV}$ then extrapolated to the
whole region. The averages are taken between $0.8<T<1.2\mbox{GeV}$,
$1.6<\omega_c<1.8\mbox{GeV}$.} \label{tdw-hqet}
\end{table}

\section{Conclusion}\label{sec5}

In conclusion, we have investigated the semileptonic decay
$\Lambda_b\rightarrow p\ell\bar\nu$ using the light-cone sum rule
method both in the full QCD and the HQET. The form factors
parameterizing the weak matrix elements of the decay are
calculated with asymptotic and QCD sum rule obtained distribution
amplitudes and used to predict the decay width and differential
distribution. For the numerical results, we find that the total
decay widths given in the full QCD are in agreement with previous
theoretical works and not inconsistent with the experimental upper
limit. However, the total decay widths with sum rule obtained
distribution amplitudes in the HQET are typically one order
smaller than those in the full QCD. What remarkable here is that
the asymptotic HQET and full QCD results are almost the same with
the fitted form factors, cf. Tables \ref{tdw-fqcd} and
\ref{tdw-hqet}. As for the physical behavior of the distribution
amplitudes, the current experimental data is not enough to
discriminate which one, between the asymptotic and the QCD sum
rule obtained, is correct in this kinematical region, because the
total decay widths corresponding to different distribution
amplitudes do not deviate from each other greatly, though the
predicted behaviors for the form factors are completely different.
Moreover, the $\mathcal{O}(\alpha_s)$ correction to the LCSR,
which amounts to the hard gluon exchange mechanism, is needed as a
complete description for the decay in the full QCD. As to the HQET
case, our results are merely preliminary and further improvements,
which including order $\bar\Lambda_{QCD}/m_Q$ corrections in the
heavy quark expansion and radiative corrections, are needed to
reconcile the existing discrepancy between the HQET and full QCD
results. Finally, it is worth remarking that the light-cone
expansion for the correlation function is not accessible at all
momentum transfers. The restriction on $q^2$ is in the region
$q^2\leq m_b^2-O(\Lambda_{\rm QCD}m_b)\approx 17 {\rm GeV}^2$.
When $q^2$ becomes large or one gets too close to the physical
states in the channel, light-cone expansion tends to break down.
So the LCSR in the large $q^2$ region may be not reliable and
cannot give much information on the physical behavior or the form
factors. The HQET case is similar, the only difference is that the
expansion parameter is $\omega$ and correspondingly the
restriction turns to be the lower limit of $\omega$, thus soft
proton will enhance the higher twist contributions.
\acknowledgments We are very grateful to C. S. Lam for his careful reading of the
manuscript and for useful suggestions. We also wish to thank A. Lenz for
communications and discussions. MQH would like to thank the McGill University for
hospitality. This work was supported in part by the National Natural Science
Foundation of China under Contract No. 10275091.
\appendix
\section*{Appendix: The nucleon distribution amplitudes with $O(x^2)$ corrections}

The nucleon distribution amplitudes are defined by matrix element
\begin{eqnarray}
&&4\langle0|\epsilon^{ijk}u_\alpha^i(a_1 x)u_\beta^j(a_2 x)u_\gamma^k(a_3
x)|P\rangle=\mathcal{S}_1MC_{\alpha\beta}(\gamma_5N)_{\gamma}+
\mathcal{S}_2M^2C_{\alpha\beta}(\rlap/x\gamma_5N)_{\gamma}\nonumber\\
&&{}+ \mathcal{P}_1M(\gamma_5C)_{\alpha\beta}N_{\gamma}+
\mathcal{P}_2M^2(\gamma_5C)_{\alpha\beta}(\rlap/xN)_{\gamma}+
(\mathcal{V}_1+\frac{x^2M^2}{4}\mathcal{V}_1^M)(\rlap/PC)_{\alpha\beta}(\gamma_5N)_{\gamma}
\nonumber\\&&{}+
\mathcal{V}_2M(\rlap/PC)_{\alpha\beta}(\rlap/x\gamma_5N)_{\gamma}+
\mathcal{V}_3M(\gamma_\mu C)_{\alpha\beta}(\gamma^\mu\gamma_5N)_{\gamma}+
\mathcal{V}_4M^2(\rlap/xC)_{\alpha\beta}(\gamma_5N)_{\gamma}\nonumber\\&&{}+
\mathcal{V}_5M^2(\gamma_\mu
C)_{\alpha\beta}(i\sigma^{\mu\nu}x_\nu\gamma_5N)_{\gamma} +
\mathcal{V}_6M^3(\rlap/xC)_{\alpha\beta}(\rlap/x\gamma_5N)_{\gamma}\nonumber\\&&{}
+(\mathcal{A}_1+\frac{x^2M^2}{4}\mathcal{A}_1^M)(\rlap/P\gamma_5
C)_{\alpha\beta}N_{\gamma}+
\mathcal{A}_2M(\rlap/P\gamma_5C)_{\alpha\beta}(\rlap/xN)_{\gamma}+
\mathcal{A}_3M(\gamma_\mu\gamma_5 C)_{\alpha\beta}(\gamma^\mu
N)_{\gamma}\nonumber\\&&{}+
\mathcal{A}_4M^2(\rlap/x\gamma_5C)_{\alpha\beta}N_{\gamma}+
\mathcal{A}_5M^2(\gamma_\mu\gamma_5 C)_{\alpha\beta}(i\sigma^{\mu\nu}x_\nu
N)_{\gamma}+ \mathcal{A}_6M^3(\rlap/x\gamma_5C)_{\alpha\beta}(\rlap/x
N)_{\gamma}\nonumber\\&&{}+(\mathcal{T}_1+\frac{x^2M^2}{4}\mathcal{T}_1^M)(P^\nu
i\sigma_{\mu\nu}C)_{\alpha\beta}(\gamma^\mu\gamma_5
N)_{\gamma}+\mathcal{T}_2M(x^\mu P^\nu i\sigma_{\mu\nu}C)_{\alpha\beta}(\gamma_5
N)_{\gamma}\nonumber\\&&{}+
\mathcal{T}_3M(\sigma_{\mu\nu}C)_{\alpha\beta}(\sigma^{\mu\nu}\gamma_5
N)_{\gamma}+
\mathcal{T}_4M(P^\nu\sigma_{\mu\nu}C)_{\alpha\beta}(\sigma^{\mu\rho}x_\rho\gamma_5
N)_{\gamma}\nonumber\\&&{}+ \mathcal{T}_5M^2(x^\nu
i\sigma_{\mu\nu}C)_{\alpha\beta}(\gamma^\mu\gamma_5 N)_{\gamma}+
\mathcal{T}_6M^2(x^\mu P^\nu i\sigma_{\mu\nu}C)_{\alpha\beta}(\rlap/x\gamma_5
N)_{\gamma}\nonumber\\&&{}+
\mathcal{T}_7M^2(\sigma_{\mu\nu}C)_{\alpha\beta}(\sigma^{\mu\nu}\rlap/x\gamma_5
N)_{\gamma}+
\mathcal{T}_8M^3(x^\nu\sigma_{\mu\nu}C)_{\alpha\beta}(\sigma^{\mu\rho}x_\rho\gamma_5
N)_{\gamma}.\label{de-def}
\end{eqnarray}
The calligraphic distribution amplitudes do not have definite twist, but can be
related to the ones with definite twist as
\begin{eqnarray}
&&\mathcal{S}_1=S_1, \hspace{0.8cm}2P\cdot x\mathcal{S}_2=S_1-S_2,\nonumber\\&&
\mathcal{P}_1=P_1, \hspace{0.8cm}2P\cdot x\mathcal{P}_2=P_1-P_2
\end{eqnarray}
for the scalar and pseudo-scalar distributions,
\begin{eqnarray}
&&\mathcal{V}_1=V_1, \hspace{2.4cm}2P\cdot x\mathcal{V}_2=V_1-V_2-V_3,
\nonumber\\&& 2\mathcal{V}_3=V_3, \hspace{2.2cm} 4P\cdot
x\mathcal{V}_4=-2V_1+V_3+V_4+2V_5,\nonumber\\&& 4P\cdot
x\mathcal{V}_5=V_4-V_3,\hspace{0.5cm} (2P\cdot
x)^2\mathcal{V}_6=-V_1+V_2+V_3+V_4+V_5-V_6
\end{eqnarray}
for the vector distributions,
\begin{eqnarray}
&&\mathcal{A}_1=A_1, \hspace{2.4cm}2P\cdot x\mathcal{A}_2=-A_1+A_2-A_3,
\nonumber\\&& 2\mathcal{A}_3=A_3, \hspace{2.2cm}4P\cdot
x\mathcal{A}_4=-2A_1-A_3-A_4+2A_5, \nonumber\\&& 4P\cdot
x\mathcal{A}_5=A_3-A_4,\hspace{0.5cm} (2P\cdot
x)^2\mathcal{A}_6=A_1-A_2+A_3+A_4-A_5+A_6
\end{eqnarray}
for the axial vector distributions, and
\begin{eqnarray}
&&\mathcal{T}_1=T_1, \hspace{3.85cm}2P\cdot x\mathcal{T}_2=T_1+T_2-2T_3,
\nonumber\\&& 2\mathcal{T}_3=T_7,\hspace{3.68cm} 4P\cdot
x\mathcal{T}_4=T_1-T_2-2T_5, \nonumber\\&& 4P\cdot x\mathcal{T}_5=-T_1+T_5+2T_8,
\hspace{0.5cm}(2P\cdot x)^2\mathcal{T}_6=2T_2-3T_3-2T_4+2T_5+2T_7+2T_8,
\nonumber\\&& 4P \cdot x\mathcal{T}_7=T_7-T_8, \hspace{1.90cm}(2P\cdot
x)^2\mathcal{T}_8=-T_1+T_2 +T_5-T_6+2T_7+2T_8
\end{eqnarray}
for the tensor distributions. Each distribution amplitudes $F=V_i, A_i, T_i, S_i,
P_i$ can be represented as
\begin{equation}
F(a_ip\cdot x)=\int \mathcal{D}x e^{-ip\cdot x\Sigma_ix_ia_i}F(x_i)\;.
\end{equation}
Those distribution amplitudes are scale dependent and can be expanded in
contributions of conformal operators. To the next-to-leading conformal spin
accuracy the expansion reads \cite{hitwist}
\begin{eqnarray}
V_1(x_i,\mu)&=&120x_1x_2x_3[\phi_3^0(\mu)+\phi_3^+(\mu)(1-3x_3)],\nonumber\\
V_2(x_i,\mu)&=&24x_1x_2[\phi_4^0(\mu)+\phi_3^+(\mu)(1-5x_3)],\nonumber\\
V_3(x_i,\mu)&=&12x_3\{\psi_4^0(\mu)(1-x_3)+\psi_4^-(\mu)[x_1^2+x_2^2-x_3(1-x_3)]
\nonumber\\&+&\psi_4^+(\mu)(1-x_3-10x_1x_2)\},\nonumber\\
V_4(x_i,\mu)&=&3\{\psi_5^0(\mu)(1-x_3)+\psi_5^-(\mu)[2x_1x_2-x_3(1-x_3)]
\nonumber\\&+&\psi_5^+(\mu)[1-x_3-2(x_1^2+x_2^2)]\},\nonumber\\
V_5(x_i,\mu)&=&6x_3[\phi_5^0(\mu)+\phi_5^+(\mu)(1-2x_3)],\nonumber\\
V_6(x_i,\mu)&=&2[\phi_6^0(\mu)+\phi_6^+(\mu)(1-3x_3)],\nonumber\\
A_1(x_i,\mu)&=&120x_1x_2x_3\phi_3^-(\mu)(x_2-x_1),\nonumber\\
A_2(x_i,\mu)&=&24x_1x_2\phi_4^-(\mu)(x_2-x_1),\nonumber\\
A_3(x_i,\mu)&=&12x_3(x_2-x_1)\{(\psi_4^0(\mu)+\psi_4^+(\mu))+\psi_4^-(\mu)(1-2x_3)
\},\nonumber\\
A_4(x_i,\mu)&=&3(x_2-x_1)\{-\psi_5^0(\mu)+\psi_5^-(\mu)x_3
+\psi_5^+(\mu)(1-2x_3)\},\nonumber\\
A_5(x_i,\mu)&=&6x_3(x_2-x_1)\phi_5^-(\mu)\nonumber\\
A_6(x_i,\mu)&=&2(x_2-x_1)\phi_6^-(\mu),\nonumber\\
T_1(x_i,\mu)&=&120x_1x_2x_3[\phi_3^0(\mu)+\frac{1}{2}(\phi_3^--\phi_3^+)(\mu)(1-3x_3)
],\nonumber\\
T_2(x_i,\mu)&=&24x_1x_2[\xi_4^0(\mu)+\xi_4^+(\mu)(1-5x_3)],\nonumber\\
T_3(x_i,\mu)&=&6x_3\{(\xi_4^0+\phi_4^0+\psi_4^0)(\mu)(1-x_3)+
(\xi_4^-+\phi_4^--\psi_4^-)(\mu)[x_1^2+x_2^2-x_3(1-x_3)]
\nonumber\\&+&(\xi_4^++\phi_4^++\psi_4^+)(\mu)(1-x_3-10x_1x_2)\},\nonumber\\
T_7(x_i,\mu)&=&6x_3\{(-\xi_4^0+\phi_4^0+\psi_4^0)(\mu)(1-x_3)+
(-\xi_4^-+\phi_4^--\psi_4^-)(\mu)[x_1^2+x_2^2-x_3(1-x_3)]
\nonumber\\&+&(-\xi_4^++\phi_4^++\psi_4^+)(\mu)(1-x_3-10x_1x_2)\},\nonumber\\
T_4(x_i,\mu)&=&\frac{3}{2}\{(\xi_5^0+\phi_5^0+\psi_5^0)(\mu)(1-x_3)+
(\xi_5^-+\phi_5^--\psi_5^-)(\mu)[2x_1x_2-x_3(1-x_3)]
\nonumber\\&+&(\xi_5^++\phi_5^++\psi_5^+)(\mu)(1-x_3-2(x_1^2+x_2^2))\},\nonumber\\
T_8(x_i,\mu)&=&\frac{3}{2}\{(-\xi_5^0+\phi_5^0+\psi_5^0)(\mu)(1-x_3)+
(-\xi_5^-+\phi_5^--\psi_5^-)(\mu)[2x_1x_2-x_3(1-x_3)]
\nonumber\\&+&(-\xi_5^++\phi_5^++\psi_5^+)(\mu)(1-x_3-2(x_1^2+x_2^2))\},\nonumber\\
T_5(x_i,\mu)&=&6x_3[\xi_5^0(\mu)+\xi_5^+(\mu)(1-2x_3)],\nonumber\\
T_6(x_i,\mu)&=&2[\phi_6^0(\mu)+\frac{1}{2}(\phi_6^--\phi_6^+)(\mu)(1-3x_3)].
\label{da-xi}
\end{eqnarray}
$V_1$, $A_1$ and $T_1$ are leading twist-$3$ distribution amplitudes; $V_2$,
$V_3$, $A_2$, $A_3$, $T_2$, $T_3$ and $T_7$ are twist-$4$; $V_4$, $V_6$, $A_4$,
$A_6$, $T_4$, $T_5$ and $T_8$ are twist-$5$; while the twist-$6$ distribution
amplitudes are $V_6$, $A_6$ and $T_6$. All the $24$ parameters involved in Eq.
(\ref{da-xi}) have been analyzed in \cite{hitwist}. It turns out that those
parameters can be expressed in terms of $8$ independent matrix elements of  local
operators. To the leading conformal spin accuracy, there are three parameters
entering
\begin{eqnarray}
\phi_3^0 = \phi_6^0 = f_N \,,\hspace{0.3cm} &\qquad& \phi_4^0 = \phi_5^0 =
\frac{1}{2} \left(\lambda_1 + f_N\right) \,,
\nonumber \\
\xi_4^0 = \xi_5^0 = \frac{1}{6} \lambda_2\,, &\qquad& \psi_4^0  = \psi_5^0 =
\frac{1}{2}\left(f_N - \lambda_1 \right)  \,. \nonumber
\end{eqnarray}
The remaining five parameters are related to the next-to-leading conformal spin
contributions
\begin{eqnarray}
\tilde\phi_3^- &=& \frac{21}{2} A_1^u,\nonumber\\
\tilde\phi_3^+ &=& \frac{7}{2} (1 - 3 V_1^d),\nonumber\\
\phi_4^- &=& \frac{5}{4} \left(\lambda_1(1- 2 f_1^d -4 f_1^u) + f_N( 2 A_1^u -
1)\right) \,,
\nonumber \\
\phi_4^+ &=& \frac{1}{4} \left( \lambda_1(3- 10 f_1^d) - f_N( 10 V_1^d -
3)\right)\,,
\nonumber \\
\psi_4^- &=& - \frac{5}{4} \left(\lambda_1(2- 7 f_1^d + f_1^u) + f_N(A_1^u + 3
V_1^d - 2)\right) \,,
\nonumber \\
\psi_4^+ &=& - \frac{1}{4} \left(\lambda_1 (- 2 + 5 f_1^d + 5 f_1^u) + f_N( 2 + 5
A_1^u - 5 V_1^d)\right)\,,
\nonumber \\
\xi_4^- &=& \frac{5}{16} \lambda_2(4- 15 f_2^d)\,,
\nonumber \\
\xi_4^+ &=& \frac{1}{16} \lambda_2 (4- 15 f_2^d)\,,\nonumber\\
\phi_5^- &=& \frac{5}{3} \left(\lambda_1(f_1^d - f_1^u) + f_N( 2 A_1^u -
1)\right) \,,
\nonumber \\
\phi_5^+ &=& - \frac{5}{6} \left(\lambda_1 (4 f_1^d - 1) + f_N( 3 + 4
V_1^d)\right)\,,
\nonumber \\
\psi_5^- &=& \frac{5}{3} \left(\lambda_1 (f_1^d - f_1^u) + f_N( 2 - A_1^u - 3
V_1^d)\right)\,,
\nonumber \\
\psi_5^+ &=& -\frac{5}{6} \left(\lambda_1 (- 1 + 2 f_1^d +2 f_1^u) + f_N( 5 + 2
A_1^u -2 V_1^d)\right)\,,
\nonumber \\
\xi_5^- &=& - \frac{5}{4} \lambda_2 f_2^d\,,
\nonumber \\
\xi_5^+ &=&  \frac{5}{36} \lambda_2 (2 - 9 f_2^d)\,,
\nonumber \\
\phi_6^- &=& \phantom{-}\frac{1}{2} \left(\lambda_1 (1- 4 f_1^d - 2 f_1^u) +
f_N(1 +  4 A_1^u )\right) \,,
\nonumber \\
\phi_6^+ &=& - \frac{1}{2}\left(\lambda_1  (1 - 2 f_1^d) + f_N ( 4 V_1^d -
1)\right)\,.
\end{eqnarray}
In Table \ref{tab-v1d} we give the asymptotic and QCD sum rule (QCDSR) obtained
numerical values for the five next-to-leading conformal spin accuracy parameters.
\begin{table}[htb]
\begin{tabular}{*{6}{@{\hspace{0.2cm}}c}}\hline\hline
&$V_1^d$&$A_1^u$&$f_1^d$&$f_2^d$&$f_1^u$\\\hline QCDSR
&$0.23\pm0.03$&$0.38\pm0.15$&$0.6\pm0.2$&$0.15\pm0.06$&$0.22\pm0.15$
\\\hline
asymptotic
&$1/3$&$0$&$3/10$&$4/15$&$1/10$\\
\hline\hline
\end{tabular}
\caption{Numerical value for the next-to-leading conformal spin
parameters.} \label{tab-v1d}
\end{table}

The twist of the order $O(x^2)$ corrections starts from twist
five, i.e., what we have write explicitly in the definition
(\ref{de-def}). Generally speaking, the derivation of the explicit
forms for those corrections are difficult. But for the special
case concerned in this article the task can be achieved using the
technique developed for the mesonic operators in
\cite{propagator}, for in our special configuration (\ref{tnuth})
the two quarks always appear at the same space-time point. What we
have done is exactly the same as that in \cite{nucleon}, so we
only present the final results, and as for the detailed procedure
it is recommended to consult the original paper. The moment
equations we obtained are
\begin{eqnarray}
\int dx_3x_3^n \mathcal{V}_1^{M(d)}(x_3) &=
&\frac{V_1^{(d)(n+2)}}{n+1} -
\frac{(V_1-V_2)^{(d)(n+2)}}{(n+1)(n+3)} -
\frac{(V_1+V_5)^{(d)(n+1)}}{(n+1)(n+3)}
\nonumber\\&+&\frac{(-2V_1+V_3+V_4+2V_5)^{(d)(n+2)}}{(n+1)(n+2)},\nonumber\\
\int dx_2x_2^n\mathcal{V}_1^{M(u)}(x_2) &=
&\frac{V_1^{(u)(n+2)}}{n+1} -
\frac{(V_1-V_2)^{(u)(n+2)}}{(n+1)(n+3)}
+\frac{(-2V_1+V_3+V_4+2V_5)^{(u)(n+2)}}{(n+1)(n+2)}
\end{eqnarray}
for the vector distributions, and
\begin{eqnarray}
\int dx_3x_3^n \mathcal{A}_1^{M(d)}(x_3) &=
&\frac{A_1^{(d)(n+2)}}{n+1} -
\frac{(A_1-A_2)^{(d)(n+2)}}{(n+1)(n+3)} -
\frac{(A_1+A_5)^{(d)(n+1)}}{(n+1)(n+3)}
\nonumber\\&+&\frac{(-2A_1-A_3-A_4+2A_5)^{(d)(n+2)}}{(n+1)(n+2)},\nonumber\\
\int dx_2x_2^n\mathcal{A}_1^{M(u)}(x_2) &=
&\frac{A_1^{(u)(n+2)}}{n+1} -
\frac{(A_1-A_2)^{(u)(n+2)}}{(n+1)(n+3)}
+\frac{(-2A_1-A_3-A_4+2A_5)^{(u)(n+2)}}{(n+1)(n+2)}
\end{eqnarray}
for the axial vector distributions. Correspondingly, the solutions
are \cite{nucleon}
\begin{eqnarray}
&& \mathcal{V}_1^{M(d)}(x_3) =\frac{x_3^2}{24}(\lambda_1
C_\lambda^d+f_N C_f^d),\nonumber\\&&
\mathcal{V}_1^{M(u)}(x_2)=\frac{x_2^2}{24}(\lambda_1
C_\lambda^u+f_N C_f^u)
\end{eqnarray}
with
\begin{eqnarray}
&&C_\lambda^d= -(1-x_3)[11+131\,x_3-169x_3^2+63x_3^3-30\,f_1^d
\,(3+11x_3-17x_3^2+7x_3^3)]-12\,(3-10\,f_1^d)\,\ln
x_3,\nonumber\\&& C_f^d= -( 1 - x_3 )
\,[1441+505x_3-3371x_3^2+3405x_3^3-1104x_3^4-24V_1^d(207-3x_3-368x_3^2+412x_3^3-138x_3^4)]
\nonumber\\&&{} - 12(73-220\,V_1^d)\,\ln x_3,\nonumber\\&&
C_\lambda^u=
-(1-x_2)^3[13-20f_1^d+3x_2+10f_1^u(1-3x_2)],\nonumber\\&& C_f^u=
(1-x_2)^3[113+495x_2-552x_2^2+10A_1^u(-1+3x_2)+2V_1^d(113-951x_2+828x_2^2)],
\end{eqnarray}
and
\begin{eqnarray}
&&\mathcal{A}_1^{M(d)}(x_3)=0,\nonumber\\&&
\mathcal{A}_1^{M(u)}(x_2)=\frac{x_2^2}{24}( 1- x_2 )^3(\lambda_1
D_\lambda^u+f_N D_f^u)
\end{eqnarray}
with
\begin{eqnarray}
&& D_\lambda^u=29-45x_2-10f_1^u(7-9x_2)-20f_1^d(5-6x_2)
,\nonumber\\&&
D_f^u=11+45x_2+10V_1^d(1-30x_2)-2A_1^u(113-951x_2+828x_2^2).
\end{eqnarray}
The tensor corrections can be obtained from the vector and axial vector
corrections by the symmetry relation between them \cite{hitwist}.



\newpage

{\bf Figure Captions}
\begin{center}
\begin{minipage}{12cm}
{\sf Fig. 1.}{\quad  Diagrammatic representation of the correlation functions
(\ref{tnu}), where the thick solid line denotes the heavy quark.}
\end{minipage}
\end{center}

\begin{center}
\begin{minipage}{12cm}
{\sf Fig. 2.}{\quad  Light-cone sum rules for the form factors
$f_1$ and $f_2$ at $q^2=0$ with parameters for the distribution
amplitudes obtained from QCD sum rules. The continuum threshold is
$s_0=39\mbox{GeV}^2$ and the heavy quark mass is
$m_b=4.8\mbox{GeV}$.}
\end{minipage}
\end{center}

\begin{center}
\begin{minipage}{12cm}
{\sf Fig. 3.}{\quad  Light-cone sum rules for the form factors $f_1$ and $f_2$.
The ``SR" in the figure denotes the result with sum rule obtained distribution
amplitudes, and the ``ASY" denotes result with asymptotic parameters. The
continuum threshold and the Borel parameter are $s_0=39\mbox{GeV}^2$,
$M_B^2=8\mbox{GeV}^2$, and the heavy quark mass is $m_b=4.8\mbox{GeV}$.}
\end{minipage}
\end{center}

\begin{center}
\begin{minipage}{12cm}
{\sf Fig. 4.}{\quad  Light-cone sum rules for the form factors
$F_1$ and $F_2$ at $\omega=2.9\mbox{GeV}$ with QCD sum rule
obtained distribution amplitudes. The continuum threshold is
$\omega_c=1.7\mbox{GeV}$ and the effective mass is
$\bar\Lambda=0.8\mbox{GeV}$.}
\end{minipage}
\end{center}

\begin{center}
\begin{minipage}{12cm}
{\sf Fig. 5.}{\quad  Light-cone sum rules for the form factors
$F_1$ and $F_2$ with $\omega_c=1.7\mbox{GeV}$, $T=1\mbox{GeV}$ and
$\bar\Lambda=0.8\mbox{GeV}$.}
\end{minipage}
\end{center}

\begin{center}
\begin{minipage}{12cm}
{\sf Fig. 6.}{\quad  Differential decay rates in the full QCD. The legends and
parameters are the same as those in Fig. 3.}
\end{minipage}
\end{center}

\begin{center}
\begin{minipage}{12cm}
{\sf Fig. 7.}{\quad  Differential decay rates in the HQET. The
legends and parameters are the same as those in Fig. 5.}
\end{minipage}
\end{center}

\newpage
\begin{figure}
\epsfxsize=6cm \centerline{\epsffile{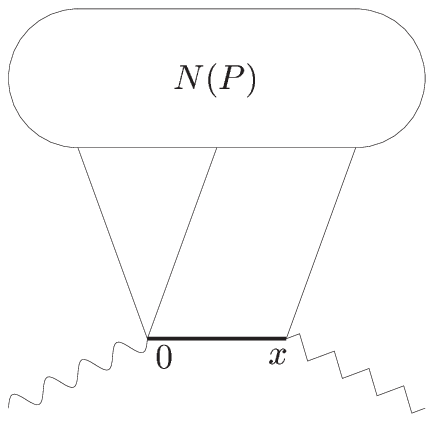}} \caption{}
\end{figure}

\begin{figure}
\begin{minipage}{7cm}
\epsfxsize=7cm \centerline{\epsffile{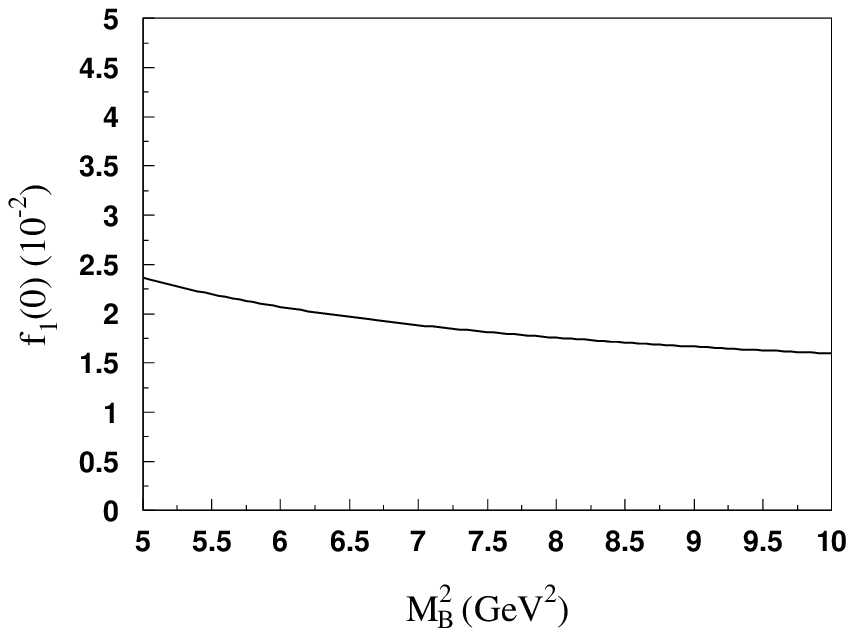}}
\end{minipage}
\hfill
\begin{minipage}{7cm}
\epsfxsize=7cm \centerline{\epsffile{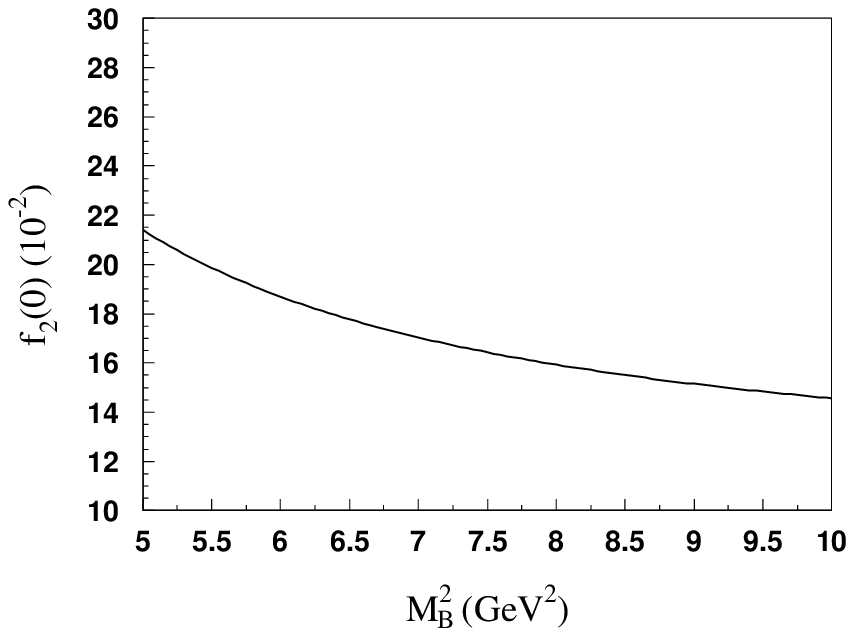}}
\end{minipage}
\caption{}
\end{figure}

\begin{figure}
\begin{minipage}{7cm}
\epsfxsize=7cm \centerline{\epsffile{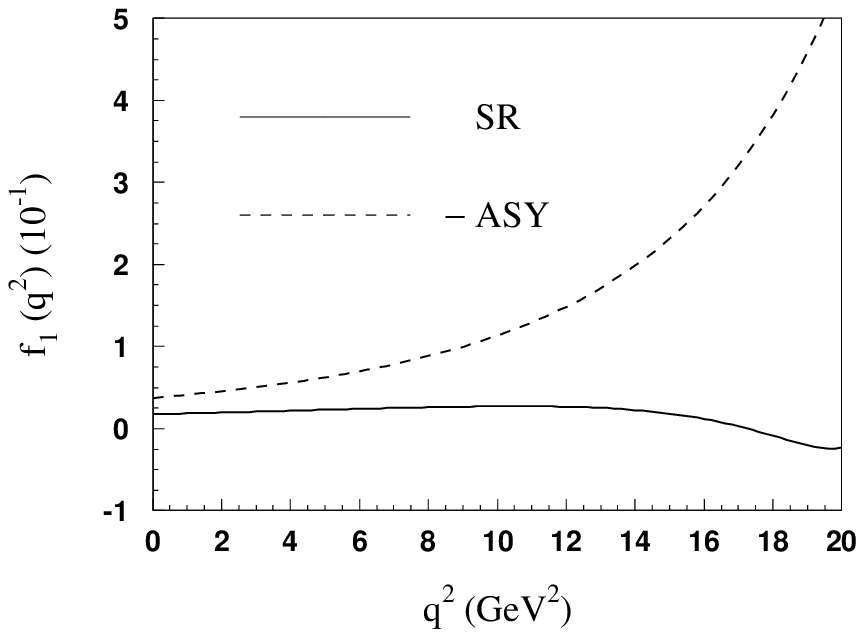}}
\end{minipage}
\hfill
\begin{minipage}{7cm}
\epsfxsize=7cm \centerline{\epsffile{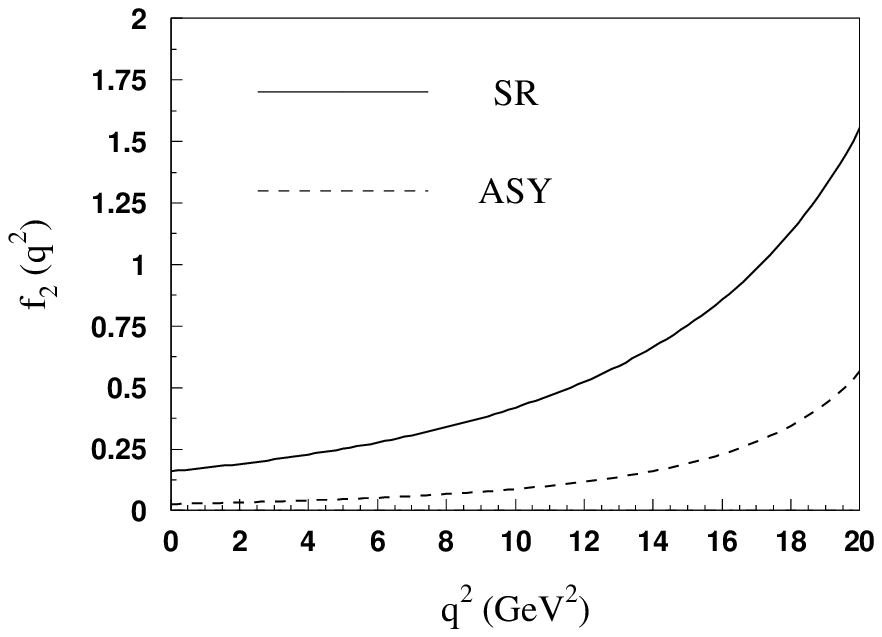}}
\end{minipage}
\caption{}
\end{figure}

\begin{figure}
\begin{minipage}{7cm}
\epsfxsize=7cm \centerline{\epsffile{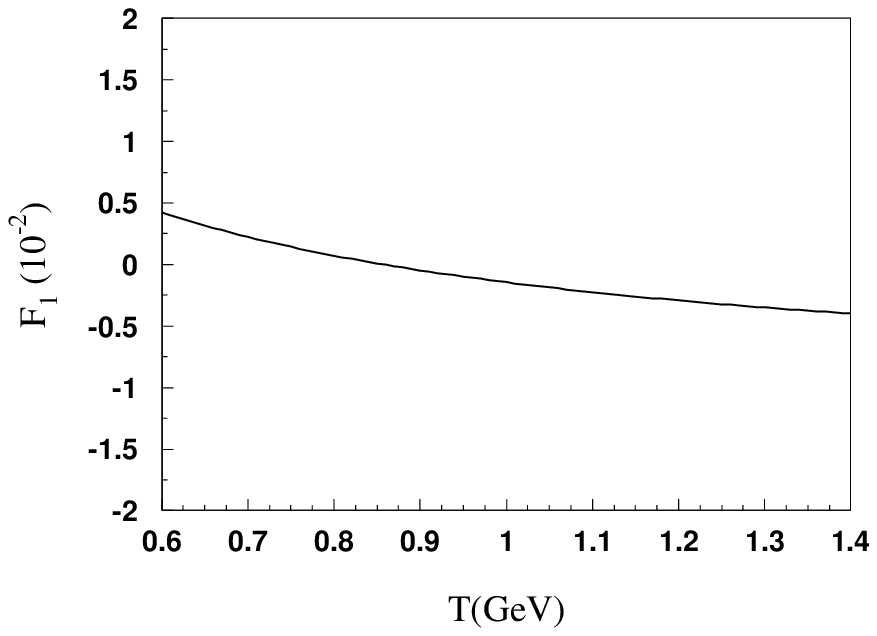}}
\end{minipage}
\hfill
\begin{minipage}{7cm}
\epsfxsize=7cm \centerline{\epsffile{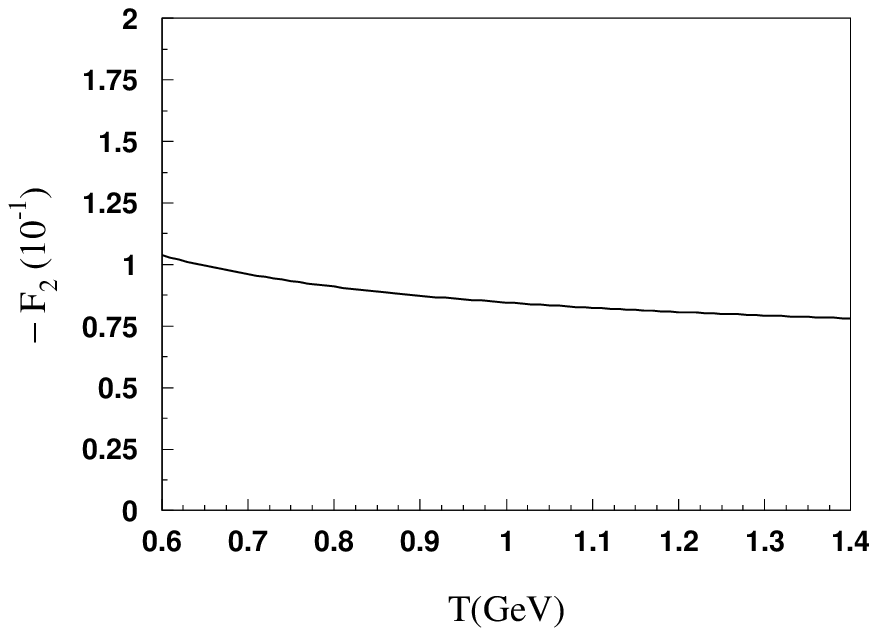}}
\end{minipage}
\caption{}
\end{figure}

\begin{figure}
\begin{minipage}{7cm}
\epsfxsize=7cm \centerline{\epsffile{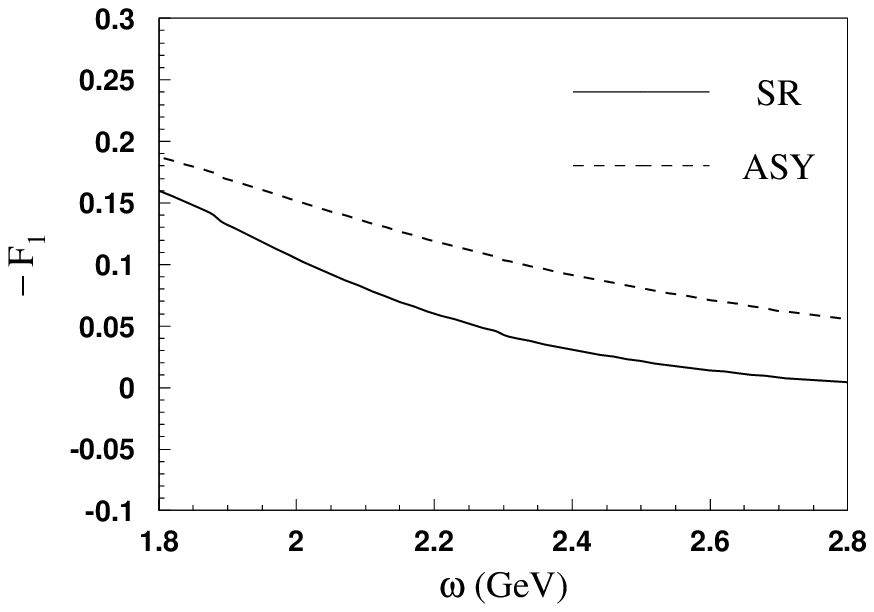}}
\end{minipage}
\hfill
\begin{minipage}{7cm}
\epsfxsize=7cm \centerline{\epsffile{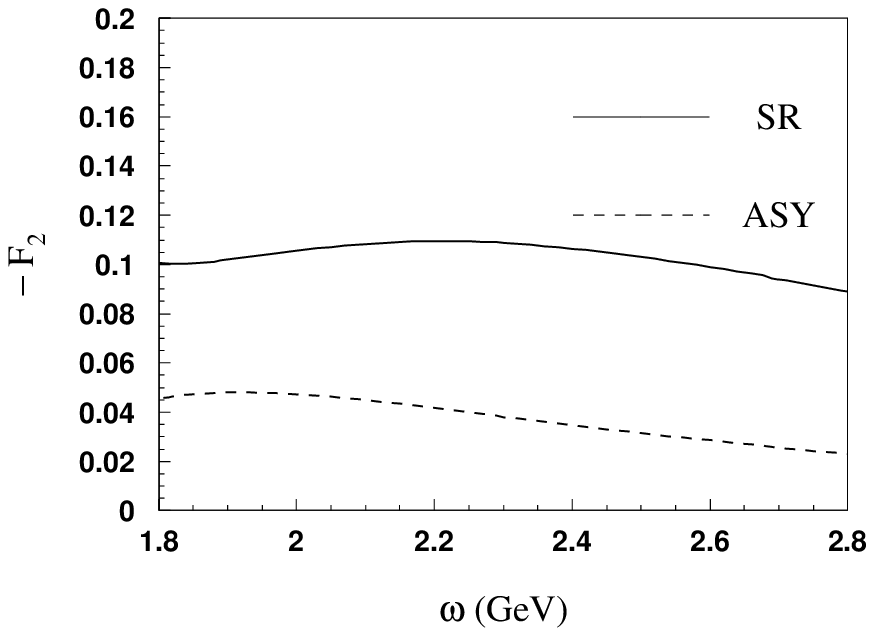}}
\end{minipage}
\caption{}
\end{figure}

\begin{figure}
\epsfxsize=10cm \centerline{\epsffile{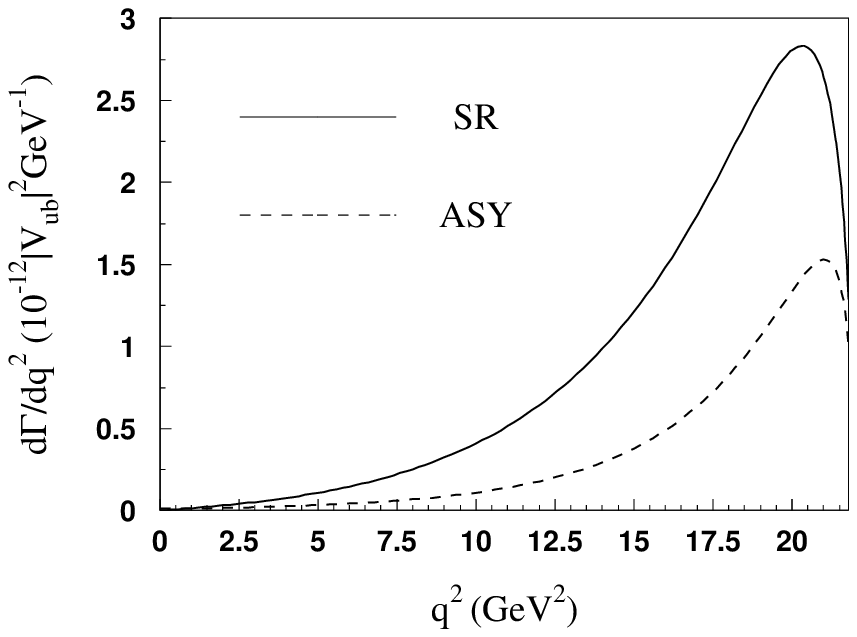}}  \caption{}
\end{figure}
\begin{figure}
\epsfxsize=10cm \centerline{\epsffile{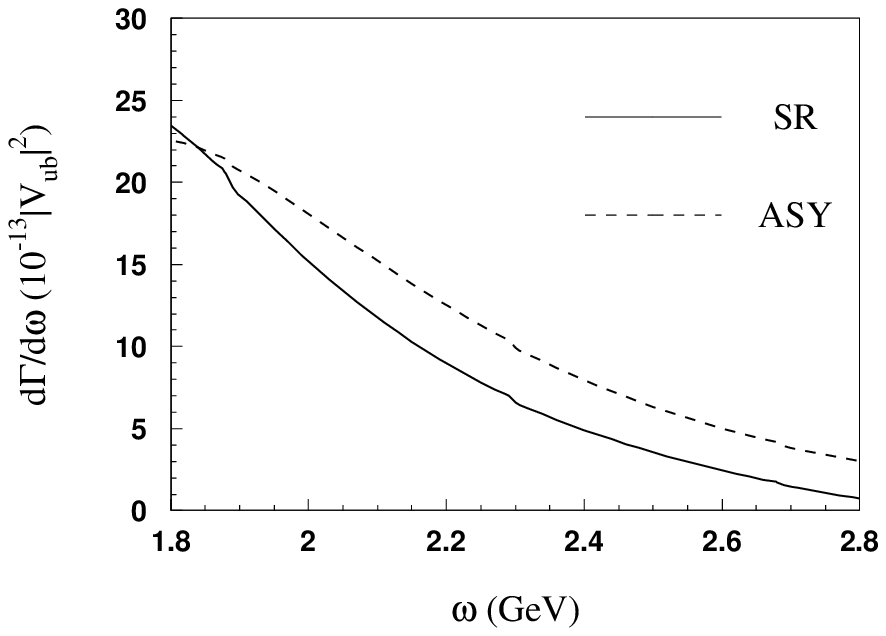}} \caption{}
\end{figure}


\begin{thebibliography}{99}
\bibitem{BBK}
I.I.\ Balitsky, V.M.\ Braun and A.V.\ Kolesnichenko, Nucl.\ Phys.\ B {\bf 312},
 509( 1989); V.L.\ Chernyak and I.R.\ Zhitnitsky, $ibid.$\ B {\bf 345}, 137 (1990).

\bibitem{LCSR}
V.~Braun, Light-Cone Sum Rules, hep-ph/9801222; V.~Braun, Recent Development of
QCD Sum Rules in Heavy Flavor Physics, hep-ph/9911206.

\bibitem{hqetsum} For a recent review on the QCD and light-cone QCD sum rule methods see:
P.~Colangelo and A.~Khodjamirian, hep-ph/0010175, published in the Boris Ioffe
Festschrift ``At the Frontier of Particle Physics/Handbook of QCD'', 1495-1576,
edited by M.~Shifman (World Scientific, Singapore, 2001).

\bibitem{svzsum} M.~A.~Shifman, A.~I.~Vainshtein and V.~I.~Zakharov, Nucl. Phys.
{\bf  B147}, 385 (1979); {\bf B147}, 448 (1979); V.~A.~Novikov, M.~A.~Shifman,
A.~I.~Vainshtein and V.~I.~Zakharov, Fortschr.\ Phys.\ {\bf 32}, 11 (1984).

\bibitem{HEP} G.~P.~Lepage and S.~J.~Brodsky, Phys. Rev. Lett. {\bf 43},
545(1979); {\bf 43}, 1625(E) (1979); G.~P.~Lepage and S.~J.~
Brodsky, Phys. Rev. D. {\bf 22}, 2157 (1980); V.~L.~Chernyak and
A.~R.~Zhitnitsky, Phys. Rept. {\bf 112}, 173 (1984).

\bibitem{apps}
V.~M.~Braun and I.~E.~Filyanov, Z. Phys. C {\bf 44}, 157 (1989); P.~Ball,
V.~M.~Braun and H.~G.~Dosch, Phys. Rev. D {\bf 44}, 3567 (1991); V.~M.~Belyaev,
A.~Khodjamirian and R.~R\"{u}ckl, Z. Phys. C {\bf 60}, 349 (1993); V.~Braun and
I.~Halperin, Phys. Lett. B {\bf 328}, 457 (1994); V.~M.~Belyaev, V.~M.~Braun,
A.~Khodjamirian and R.~R\"{u}ckl, Phys. Rev. D {\bf 51}, 6177 (1995); P.~Ball and
V.~M.~Braun, $ibid.$ D {\bf 58}, 094016 (1998); A.~ Khodjamirian, Nucl. Phys.
{\bf B605}, 558 (2001).

\bibitem{Btopi}
E.~Bagan, P.~Ball and V.M.~Braun, Phys.\ Lett.\ B {\bf 417}, 154 (1998);
A.~Khodjamirian, R.~Ruckl and C.W.~Winhart, Phys.\ Rev.\ D {\bf 58}, 054013
(1998); A.~Khodjamirian {\it et al.}, $ibid.$ D {\bf 62}, 114002 (2000); P.~Ball,
JHEP {\bf 9809}, 005 (1998); P.~Ball and R.~Zwicky, $ibid.$ {\bf 0110}, 019
(2001).

\bibitem{app-h}
A.~Ali, V.~M.~Braun, and H.~Simma, Z. Phys. C {\bf 63}, 437 (1994); P.~Ball and
V.~M.~Braun, Phys. Rev. D {\bf 55}, 5561 (1997).

\bibitem{HQET}  B.~Grinstein, Nucl. Phys. {\bf B339}, 253 (1990);
E.~Eichten and B.~Hill, Phys. Lett. B {\bf  234}, 511 (1990); A.~F.~Falk,
H.~Georgi, B.~Grinstein, and M.~B.~Wise, Nucl. Phys. {\bf B343}, 1 (1990).

\bibitem{review} M. Neubert, Phys. Rept. {\bf 245}, 259 (1994).

\bibitem{app-hqet}
Y.~B.~Dai and S.~L.~Zhu, Phys.\ Rev.\ D {\bf 58}, 074009 (1998); S.~L.~Zhu and
Y.~B.~Dai, Phys.\ Lett.\ B {\bf 429}, 72 (1998); S.~L.~Zhu and Y.~B.~Dai, Phys.\
Rev.\ D {\bf 59}, 114015 (1999); W. Y. Wang and Y. L. Wu, Phys.\ Lett.\ B {\bf
515}, 57 (2001); J.~G.~K\"{o}rner, C.~Liu and C.~T.~Yan, Phys.\ Rev.\ D {\bf 66},
076007 (2002).

\bibitem{hitwist}
V.~M.~Braun, R.~J.~Fries, N.~Mahnke, and E.~Stein, Nucl. Phys. {\bf B589}, 381
(2000); {\bf B607}, 433(E)(2001).

\bibitem{result1}
C.~S.~Huang, C.~F.~Qiao and H.~G.~Yan, Phys. Lett. B {\bf 437}, 403 (1998);
R.~S.~Marques de Carvalho, F.~S.~Navarra, M.~Nielsen, E.~Ferreira and
H.~G.~Dosch, Phys.\ Rev.\ D {\bf 60}, 034009 (1999).

\bibitem{result2}
A.~Datta, Semi-Leptonic Decays of $\Lambda_c$ and $\Lambda_b$ Baryons involving
Heavy to Light Transitions and the determination of $V_{ub}$, hep-ph/9504429;
M.~A.~Ivanov, V.~E.~Lyubovitskij, J.~G.~Korner and P.~Kroll, Phys.\ Rev.\ D {\bf
56}, 348 (1997);

\bibitem{pqcd}
W. Loinaz and R. Akhoury, Phys.\ Rev.\ D {\bf 53}, 1416 (1996); H.~H.~Shih,
S.~C.~Lee and H.~n.~Li, $ibid.$ D {\bf 59}, 094014 (1999).

\bibitem{current} B.~L.~Ioffe, Nucl. Phys. {\bf  B188}, 317 (1981);E: {\bf  B191}, 591
(1981); Z. Phys. C {\bf  18}, 67 (1983); V.~M.~Belyaev and B.~Yu.~Blok, Z. Phys.
C {\bf  30}, 151 (1983); M. A. Ivanov {\it et al.}, Phys. Rev. D {\bf 61}, 114010
(2000); D.~W.~Wang and M.~Q.~Huang, $ibid.$ D {\bf 67}, 074025 (2003).

\bibitem{nucleon}
V.~M.~Braun, A.~Lenz, N.~Mahnke, and E.~Stein, phys. Rev. D {\bf
65}, 074011 (2002); A.~Lenz, M.~Wittmann and E.~Stein, Phys.\
Lett.\ B {\bf 581}, 199 (2004).

\bibitem{wf}
V.~L.~Chernyak and I.~R.~Zhitnitsky, Nucl. Phys. {\bf B246}, 52 (1984);
I.~D.~King and C.~T.~Sachrajda, $ibid.$ {\bf B279}, 785 (1987);  V.~L.~Chernyak,
A.~A.~Ogloblin and I.~R.~Zhitnitsky, Sov. J. Nucl. Phys. {\bf 48}, 536 (1988); Z.
Phys. C {\bf 42}, 583 (1989).

\bibitem{propagator}
I.~I.~Balitsky and V.~M.~Braun, Nucl. Phys. {\bf B311}, 541
(1989).

\bibitem{multi-part}
M.~Diehl, T.~Feldmann, R.~Jakob, and P.~Kroll, Eur. Phys. J. C
{\bf 8}, 409 (1999).

\bibitem{pion}
V.~M.~Braun, A.~Khodjamiran, and M.~Maul, Phys. Rev. D {\bf 61},
073004 (2000).

\bibitem{H-limit}
E.~Bagan, P.~Ball, V.~M.~Braun and H.~G.~Dosch, Phys.\ Lett.\ B
{\bf 278}, 457 (1992).

\bibitem{hform}
T.~Mannel, W.~Roberts, and Z.~Ryzak, Nucl. Phys. {\bf B355}, 38
(1991).

\bibitem{mb-lb}
E.~V.~Shuryak, Nucl.\ Phys.\ B {\bf 198}, 83 (1982); A.~G.~Grozin and
O.~I.~Yakovlev, Phys.\ Lett.\ B {\bf 285}, 254 (1992); Y.~B.~Dai, C.~S.~Huang,
C.~Liu and C.~D.~Lu, $ibid.$ B {\bf 371}, 99 (1996); D.~W.~Wang, M.~Q.~Huang and
C.~Z.~Li, Phys.\ Rev.\ D {\bf 65}, 094036 (2002).

\bibitem{excited-mass}
D.~W.~Wang and M.~Q.~Huang, Phys.\ Rev.\ D {\bf 68}, 034019
(2003).

\bibitem{PDG2002}
K.~Hagiwara {\it et al.}, Particle Data Group, Phys.\ Rev.\ D {\bf
66}, 010001 (2002).

\end{thebibliography}
\end{document}